\documentclass[12pt]{article} %
\usepackage{latexsym}
\usepackage{amsfonts}
\usepackage{amsmath}
\usepackage{graphicx}
\usepackage{amssymb}

\numberwithin{equation}{section}

\def\tA{{\tt A}}

\def\tb{{\tt b}}
\def\tB{{\tt B}}

\def\tc{{\tt c}}
\def\td{{\tt r}}

\def\tth{{\tt h}}

\def\tp{{\tt p}}
\def\tP{{\tt P}}
\def\otP{{\overline\tP}}
\def\tR{{\tt R}}
\def\ttr{{\tt r}}
\def\tS{{\tt S}}
\def\ts{{\tt s}}
\def\tT{{\tt T}}
\def\ttt {{\tt t}}

\def\tU{{\tt U}}
\def\tu{{\tt u}}

\def\tw{{\tt w}}
\def\tx{{\tt x}}
\def\ty{{\tt y}}
\def\t0{{\tt 0}}

\def\eps{{\epsilon}}

\def\orho{{\overline\rho}}

\def\om{\omega}
\def\oom{\overline\omega}

\def\bom{{\mbox{\boldmath$\omega$}}}
\def\obom{\overline\bom}

\def\0bom{{\bom}^0}
\def\0obom{{\obom}^0}

\def\0nbom{{\bom}_{n,0}}
\def\n*bom{{\bom}^*_{(n)}}
\def\wt{\widetilde}

\def\oom{\overline\om}

\def\Om{\Omega}
\def\oOm{\overline\Om}

\def\bOm{\mbox{\boldmath${\Om}$}}

\def\Lam{\Lambda}
\def\Lamc{\Lam^{\rm c}}
\def\oLam{\overline\Lam}
\def\oLamc{\oLam^{\,\rm c}}

\def\oL{\overline L}

\def\fB{\mathfrak B}

\def\fF{\mathfrak F}

\def\fK{\mathfrak K}
\def\fT{\mathfrak T}

\def\fW{\mathfrak W}

\def\rd{\rm d}
\def\rmm{\rm m}

\def\cC{\mathcal C}
\def\cD{\mathcal D}

\def\cG{\mathcal G}
\def\cH{\mathcal H}

\def\cL{\mathcal L}

\def\cV{\mathcal V}
\def\cW{\mathcal W}

\def\bbP{\mathbb P}

\def\bbR{\mathbb R}

\def\bfI{\textbf I}

\def\bx{\mathbf x}

\def\ux{\underline x}

\def\by{\mathbf y}

\def\diy{\displaystyle}
\def\ov{\overline}
\def\ovl{{\ov l}}

\def\ux{\underline x}

\def\om{\omega}
\def\oom{\ov\om}

\def\rJ{{\rm J}}
\def\rd{{\rm d}}
\def\rR{{\rm R}}

\begin{document}
\title{Shift-invariance for FK-DLR states\\
of a 2D quantum bose-gas}

\author{\bf Y. Suhov$^{1}$, M. Kelbert$^{2}$, I. Stuhl$^{3}$ }
\vspace{1mm}\maketitle {\footnotesize

\noindent $^1$ Statistical Laboratory, DPMMS, University of Cambridge, UK;\\
Department of Statistics/IME,  University of S\~ao Paulo, Brazil;\\
IITP, RAS, Moscow, Russia\\
E-mail: yms@statslab.cam.ac.uk

\noindent $^2$ Department of Mathematics, Swansea University, UK;\\
Department of Statistics/IME, University of S\~ao Paulo, Brazil\\
E-mail: M.Kelbert@swansea.ac.uk

\noindent $^3$ University of Debrecen, Hungary;\\
Department of Mathematics/IME, University of S\~ao Paulo, Brazil\\
E-mail: izabella@ime.usp.br}

\begin{abstract}
This paper continues the work \cite{SK} and focuses on infinite-volume bosonic
states for a quantum  system (a quantum gas) in a plane $\bbR^2$.
We work under similar assumptions upon the form of local
Hamiltonians and the type of the (pair) interaction potential as
in \cite{SK}. The result of the paper is that any infinite-volume
FK-DLR functional corresponding to the Hamiltonians is
shift-invariant, regardless of whether this functional is unique or not.
\vskip 1 truecm

\vskip.1truecm

\end{abstract}
\vskip 1 truecm

\centerline{\large\bf{1. Introduction: FK-DLR states of quantum systems}}
\vskip 1 truecm

This work continues \cite{SK} (and earlier works \cite{KS1}, \cite{KS2} and
\cite{KSY2}). The reference to \cite{SK} are marked by the Roman number {\bf I}:
Eqn (1.1.19.{\bf I}), Theorem 1.1.{\bf I}, Section 2.3.{\bf I} and so on. In this paper we
provide a justification of the notion of an FK-DLR
(Feynman--Kac--Dobrushin--Lanford--Ruelle) state of a quantum system
in an infinite volume (more generally, an FK-DLR functional of the quasi-local C$^*$-algebra). 
The result of the present paper is that,
for a quantum Bose-gas on a plane $\bbR^2$, any FK-DLR state is shift-invariant.
This line of results takes its origin in \cite{FP}, \cite{P}; we want to stress that
a particular impact upon the present work was made by Refs \cite{R1}--\cite{R3}
(more credit will be given in due course).

We follow the background used in Sects 1.{\bfI} -- 3.{\bfI}, in a specific situation where
the dimension $d=2$. Accordingly, $\Lam=\Lam_L$ and $\Lam_0$ stand for squares
$[-L,L]^{\times 2}\subset\bbR^2$ and $[-L_0,L_0]^{\times 2}\subset\bbR^2$, or -- more generally --
$[-L_0+\tc^1,L_0+\tc^1]\times [-L_0+\tc^2,L_0+\tc^2]$, where $c=(\tc^1,\tc^2)\in\bbR^2$
and $\Lam \supset \Lam_0$. As in \cite{SK}, we
denote by $z$ and $\beta$ the standard thermodynamical variables of the fugacity
and the inverse temperature. The notions of a quantum $n$-particle Hamiltonian
$H_{n,\Lam}$ and the Gibbs state $\varphi_{z,\beta ,\Lam}$ in $\Lam$ are introduced as
in Sects 1.1.{\bf I} (see Eqns (1.1.1.{\bf I}) -- (1.1.25.{\bf I})). We also follow the
conditions upon the two-body potential $V:\,[\td ,\infty )\to\bbR$ imposed in \cite{SK}.
(Here $\td\in (0,\infty )$ is the
hard-core diameter, and we formally set
$V(r)=+\infty$ for $0\leq r<\td$, conforming with the hard-core condition.)
Moreover, we use the corresponding notation:
cf. Eqns (1.1.3.{\bf I})--(1.1.5.{\bf I}), (1.1.19.{\bf I}) and (1.2.9.{\bfI}).
For the reader's convenience,
we reproduce these conditions (and assume that they are valid throughout the paper):
$$V(r)=0\;\hbox{ for }\;r\geq\tR\;\hbox{ where }\tR\in (\td,\infty ),\eqno (1.1)$$
$$-{\ov V}=\min\,\big[V(r):\;\td\leq r\leq \tR\big],\eqno (1.2)$$
with ${\ov V}=0$ for $V\geq 0$,
$${\ov V}^{\,(1)}=\max\,\big[\left| V'(r)\right|: \td\leq r\leq\tR\big],\;\;
{\ov V}^{\,(2)}=\max\,\big[\left| V''(r)\right|: \td\leq r\leq\tR\big],\eqno (1.3)$$
and
$$\orho:=z\exp\,(4\beta {\ov V}\tR^2/\td^2)<1.\eqno (1.4)$$

The result from \cite{SK} implies that for $z>0$ and $\beta >0$ satisfying the bound
(1.4), the family of Gibbs states $\{\varphi_{z,\beta ,\Lam}\}$ is compact and has limiting points
as $\Lam\nearrow\bbR^2$. Moreover,
the family of Gibbs states $\{\varphi_{z,\beta ,\Lam |\bx (\Lamc )}\}$ is compact where
$\varphi_{z,\beta ,\Lam |\bx (\Lamc )}$ is the Gibbs state in an external
potential field generated by a `classical' configuration $\bx (\Lamc )\subset \Lamc$
satisfying (1.1.20.{\bf I}). see Theorem 1.1. The limiting points for families
$\{\varphi_{z,\beta ,\Lam}\}$ and $\{\varphi_{z,\beta ,\Lam |\bx (\Lamc )}\}$ yield states
of the quasi-local C$^*$-algebra $\fB$; see (1.2.5.{\bf I}). Such states describe
possible `thermodynamic phases' of the quantum Bose-gas in an infinite volume.
A theory proposed in \cite{SK} goes
a step further: we establish that any such limit-point state $\varphi$ has a particular
structure where the operators $R^{\Lam_0}$ yielding the (limiting) density matrices
are constructed via an FK representation.

More precisely, the integral kernels
$F^{\Lam_0}(\bx_0,\by_0)$ determining the density matrices $R^{\Lam_0}$ are written as
integrals over spaces of so-called path and loop configurations;
cf. Sect 2.{\bf I}. An important r\^ole in these formulas is played
by a probability measure (or probability measures) $\mu$ on $\cW^*(\bbR^2)$, the space
of loop configurations (LCs)
in the plane $\bbR^2$. In a natural sense, the correspondence between a functional and a
measure is one-to-one.
Such a measure $\mu$ was called an FK-DLR probability measure (PM)
and emerged as a limiting point for the family of similar measures in finite volumes $\Lam$ as
$\Lam\nearrow\bbR^2$. The set of FK-DLR PMs is denoted by $\fK=\fK (z,\beta )$; in
a probabilistic terminology these measures are examples of random marked point
processes (RMPPs) with marks represented by loops.
Accordingly, a class of states $\fF_+=\fF_+(z,\beta )$ was introduced,
called FK-DLR states, together with its enlargement, $\fF=\fF (z,\beta )\supset\fF_+$, giving a
class of FK-DLR functionals on $\fB$. See Definitions 2.4.{\bf I}--2.7.{\bf I}.
We stated a result, Theorem 1.2.{\bf I}, and its generalization, Theorem 2.2.{\bfI},
claiming that any functional from class $\fF$ is shift-invariant. For reader's convenience, we
repeat here the statements of the latter. 

\medskip

The results of this paper are summarised in the following two theorems. 
\medskip

{\it The Fock spaces
$\cH (\Lam_0)$ and $\cH ({\tS}(s)\Lam_0)$ (cf. {\rm{(1.1.12.{\bfI})}}) are related through a pair 
of mutually inverse shift isomorphisms of Fock spaces
$${\tU}^{\Lam_0}(s):\;\cH (\Lam_0)\to\cH ({\tS}(s)\Lam_0)\hbox{ and }
{\tU}^{\Lam_0}(-s):\; \cH ({\tS}(s)\Lam_0)\to\cH (\Lam_0).$$
With the shift isometry ${\tS}({\ts}):\bbR^2\to\bbR^2$:
$${\tS}(s)y=y+s,\;\;y\in\bbR^2,$$
and for the image ${\tS}(s)\Lam_0$ of $\Lam_0$:
$$\begin{array}{l}{\tS}(s)\Lam_0
=\left[{\tb}^1+{\ts}^1-L^0,{\tb}^1+{\ts}^1+L^0\right]\\
\qquad\qquad\qquad\qquad\times
\left[{\tb}^2+{\ts}^2-L^0,{\tb}^2+{\ts}^2+L^0\right].\end{array}$$
The isomorphisms ${\tU}^{\Lam_0}(s)$ and
${\tU}^{\Lam_0}(-s)$ are given by
$$\begin{array}{c}\left({\tU}^{\Lam_0}(s)\phi_n\right) (\ux_1^n)
=\phi_n ({\tS}(-s)\ux_1^n),\\
\left({\tU}^{\Lam_0}(-s)\phi_n\right) (\ux_1^n)=
\phi_n ({\tS}(s)\ux_1^n),\end{array}
\;\;\ux_1^n\in\left(\Lam_0\right)^n,$$
where $\phi_n\in
{\rm L}_2^{\rm{sym}}\left((\Lam_0)^n\right)$, $n=0,1,\ldots $.}
Cf. Eqns (1.2.6.{\bfI})--(1.2.8.{\bfI}).

\medskip

\medskip

{\bf Theorem 1.1.} (cf. Theorem 2.2.{\bf I}) {\sl Assuming conditions {\rm{(1.1)--(1.4)}}, let $\mu$
be a probability measure
from $\fK (z,\beta )$. Then the corresponding FK-DLR functional $\varphi_{\mu}\in\fF(z,\beta )$
is shift-invariant: for any square $\Lam_0\subset\bbR^2$, vector $s\in\bbR^2$ and
operator $A\in\fB (\Lam_0)$,
$$\varphi_{\mu}(U^{\tS (s)\Lam_0}(-s)AU^{\Lam_0})=\varphi_{\mu}(A).$$
In terms of the corresponding infinite-volume reduced density matrices} $R^{\Lam_0}$:
$$R^{\tS (s)\Lam_0} ={\tU}^{\Lam_0}(s)R^{\Lam_0}{\tU}^{\tS (s)\Lam_0}(-s).$$

\medskip

\medskip

In view of formulas (2.3.5.{\bf I})--(2.3.7.{\bf I}) relating an FK-DLR functional
$\varphi$ to an FK-DLR PM $\mu$, it suffices to verify

\medskip

\medskip

{\bf Theorem 1.2.} {\sl Any FK-DLR PM $\mu$ is translation invariant:
for any $s=({\ts}^1,{\ts}^2)\in\bbR^2$, square $\Lam_0=[-L_0,L_0]^{\times 2}$  and
event $\cD\in\cW^* (\bbR^2)$ localised in $\Lam_0$ (i.e., belonging to a sigma-algebra
$\fW (\Lam_0)$; cf Definition {\rm{2.4.{\bfI}}}),
$$\mu (\tS (s)\cD)=\mu (\cD).$$
Here $\tS (s)\cD$ stands for the shifted event localised in the shifted square\\
$\tS (s)\Lam_0=[-L_0+\ts^1,\ts^1+L_0]\times [-L_0+\ts^2,\ts^2+L_0]$.}

\vskip 1 truecm

\centerline{\large{\bf 2. Proof of Theorem 1.2: a tuned-shift argument}}
\vskip 1 truecm

In what follows we use the terminology and the system of notation from Sect 1.{\bfI} and 2.{\bfI}. The
proof of Theorem 1.2 is based on a modification of an argument
developed in \cite{R1}--\cite{R3}. We want to stress that the paper
\cite{R3} treating some classes of (Gibbsian) RMPPs does not cover our situation
because a number of the assumptions used in \cite{R3} are (unfortunately) not fulfilled here.
Specifically, the condition (2.2) from
\cite{R3} does not hold in our situation, as well as conditions specifying
what is called a bpsi-function on p. 704 of \cite{R3}.\footnote{In short, the paper \cite{R3}
employs an approach based on sup-norm conditions whereas the situation under consideration
in this paper requires the use of integral-type norms. A crucial fact is that a Jacobian
emerging in the course of the construction has the form (3.23) suitable for our purposes.}
The aforementioned modification
demands that we use (and inspect) the construction from \cite{R2} for
classical configurations (CCs) arising as $\ttt$-sections of LCs
at a given time point $\ttt\in [0,\beta]$.

Because the argument in the proof does not depend on the direction of the vector $s$,
we will assume that $s=(\ts,0)$ lies along the horizontal axis. Also, due to the group
property, we can assume that $\ts\in (0,1/2)$. By using constructions developed in
\cite{G} and \cite{R2}--\cite{R3}, the assertion of Theorem 1.2 can be deduced from

\medskip

{\bf Theorem 2.1.} {\sl Let $\mu$ be an FK-DLR PM, $\Lam_0$ be a square $[-L_0,L_0]^{\times 2}$
and an event $\cD\subset\cW^*_\td(\bbR^2)$ be given, localized in $\Lam_0$:
$\cD\in \fW(\Lam_0)$. Then
$$\mu (\tS(s)\cD)+\mu (\tS(-s)\cD)-2\mu (\cD)\geq 0.\eqno (2.1)$$}
\medskip

For the proof of Theorem 2.1 we employ a strategy essentially mimicking the one from 
\cite{R1}--\cite{R3}, particularly \cite{R2}. Consequently, we will follow the scheme from \cite{R2} 
rather closely, although, as was said earlier, we introduce considerable alterations. 
{\it We introduce the functionals $K$ and $L$ for path and loop configurations :
$$K(\oOm^*)=\sum_{\oom^*\in\oOm^*} k(\oom^*),\;\;K(\Om^*)=\sum_{\om^*\in\Om^*} k(\om^*),
\;\;L(\Om^*)=\prod\limits_{\om^*\in\Om^*} k(\om^*).
$$}
For a given (large) $L$ we introduce the square
$$\Lam =[-L,L]\times [-L,L]\supset\Lam_0\eqno (2.2)$$
and write the terms $\mu (\tS(\pm s)\cD)$ and $\mu (\cD)$ as integrals of conditional
expectations relative to the sigma-algebra $\fW (\Lamc )$:
$$\begin{array}{l}\diy\int_{\cW^*_\td(\bbR^2)}\mu (\rd\bOm^*_{\Lamc})
{\mathbf 1}\Big(\bOm^*_{\Lamc}\in\cW_\td\left(\Lamc \right)\Big)\\
\qquad\diy\times\int_{\cW^*_\td(\Lam )}
\rd\Om^*_{\Lam}{\mathbf 1}\Big(\Om^*_\Lam\in\tS (\pm s)\cD\Big)\frac{z^{K(\Om^*_\Lam)}}{L(\Om^*_\Lam)}
\exp\,\big[-h(\Om^*_{\Lam }|\bOm^*_{\Lamc})\,\big]
\end{array}\eqno (2.3)$$
(the case of $\mu (\cD)$ is recovered at $s=0$, with $\tS (0)={\rm{Id}}$.)

Furthermore, again as in \cite{R1}--\cite{R3}, we employ maps
$\tT^{\,\pm}_L=\tT^{\,\pm}_{L,L_0}(s):\cW^*(\bbR^2)$
$\to \cW^*_\td(\bbR^2)$.\footnote{The symbol used in  \cite{R1}--\cite{R3} is $\fT$ instead
of $\tT$. The idea of using maps $\tT^{\,\pm}_L$ goes back to
\cite{FP} and \cite{P}.} These are applied to the concatenated loop
configuration (LC) $\Om^*_{\Lam}\vee\bOm^*_{\Lamc}$ in the expressions
from Eqn (4.3), in the the corresponding
case of shift $\tS (\pm s)$. Important
properties of maps $\tT^{\,\pm}_L$ are:
\medskip

(i) The maps $(\Om^*_{\Lam},\bOm^*_{\Lamc})
\mapsto \tT^{\,\pm}_L(\Om^*_{\Lam},\bOm^*_{\Lamc})$ are one-to-one, and
a number of `nice' properties hold true when the LC
$\Om^*_{\Lam}\vee\bOm^*_{\Lamc}$ lies in a `good' set
$\cG_L\subset\cW^*_\td(\bbR^2)$. (Viz., for $\Om^*_{\Lam}\vee\bOm^*_{\Lamc}\in \cG_L$
the loops from $\Om^*_\Lam\cap\cW^*_\td(\Lam_0)$
will not interact with loops from $\bOm^*_{\Lamc}$.) The set $\cG_L$
carries asymptotically a full measure as $L\to\infty$. See below.
\medskip

(ii) For a `good'  LC $\bOm^*=
\bOm^*_{\Lam}\vee\bOm^*_{\Lamc}\in \cG_L$ over $\bbR^2$, the `external' part
$\bOm^*_{\Lamc}$ is preserved under $\tT^{\,\pm}_n$.
In other words, the maps are non-trivial only on the part $\bOm^*_{\Lam }$
(although the way $\bOm^*_{\Lam}$ is transformed depends upon $\bOm^*_{\Lamc}$
(and on $\bOm^*_\Lam$, of course)).
For that reason we will often address $\tT^{\,\pm}_L$ as a `tuned' shift
$\bOm^*\mapsto{\wt\bOm^*} =
(\tT^\pm_L\bOm^*_\Lam)\vee\bOm^*_{\Lamc}$ or, dealing with a pair
$(\Om^*_\Lam,\bOm^*_{\Lamc})\in\cW^*(\Lam,\Lamc)$,
$$\Om^*_{\Lam}\mapsto{\wt\Om^*}_\Lam
=\tT^{\,\pm}_L\Om^*_\Lam\in\cW^*_\td(\Lam ).\eqno (2.4)$$
With this agreement:
\medskip

(iii) The transformation (2.4) preserves the cardinality: $\sharp\,\Om^*_{\Lam}=\sharp\,{\wt\Om^*}_\Lam$ 
and transforms a loop $\om^*\in\Om^*_\Lam$ as $\om^*\mapsto {\wt\om^*}$ where
$\;k({\wt\om^*})=k(\om^*)$. Consequently, functionals $K$ and $L$ are preserved:
$K({\wt\Om^*})=K(\Om^*)$ and $L({\wt\Om^*})=L(\Om^*)$. Next,
for all $\ttt\in [0,k(\om^*)\beta ]$, point ${\wt\om^*}(\ttt )\in\bbR^2$ is obtained as a `tuned shift'
$${\wt\om^*}(\ttt ) =\om^*(\ttt )\pm s\tR^\pm_L \Big[\om^*;\ttt;\{\Om^*_{\Lam}\}(\ttt )\cup\{
\bOm^*_{\Lamc}\}(\ttt )\Big];\eqno (2.5)$$
see below. We stress that the argument of function $\tR^\pm_L$ consists of a loop
$\om^*\in\cW^{\,*}_\td$,
a time point $\ttt\in [0,k(\om^*)\beta]$ and the $\ttt$-section $\{\Om^*_{\Lam}\}(\ttt )\cup\{
\bOm^*_{\Lamc}\}(\ttt )=\{\Om^*_{\Lam}\vee
\bOm^*_{\Lamc}\}(\ttt )\in\cC_\td(\bbR^2)$ of an LC
$\Om^*_{\Lam}\vee\bOm^*_{\Lamc}$. 
Here, as in \cite{SK}, $\cC (\bbR^2 )$ stands for the collection of 
finite or countable (unordered) subsets $\bx\subset\bbR^2$ (including the empty set) 
and $\cC_\td(\bbR^2)\subset\cC (\bbR^2 )$ for the collection of  subsets $\bx$ with 
$$\min\;\Big[|x -x'|:\;x,x'\in\bx ,\;x\neq x' \Big]\geq \td.$$

\medskip

(iv) For simplicity, let us omit henceforth the symbols $\pm$ whenever possible.
The value $\tw_L \Big[\om^*;\ttt;\{\Om^*_{\Lam}\}(\ttt )\cup\{
\bOm^*_{\Lamc}\}(\ttt )\Big]$ is a non-negative number. Moreover, when
$\Om^*_{\Lam}\vee\bOm^*_{\Lamc}\in \cG_L$ then for
$\om^*\in\Om^*_\Lam\cap\cW^*_\td(\Lam_0)$ and $0\leq \ttt\leq k(\om^*)\beta$,
$$\tw_L \Big[\om^*;\ttt;\{\Om^*_{\Lam}\}(\ttt )\cup\{
\bOm^*_{\Lamc}\}(\ttt )\Big]\equiv 1,\;\;0\leq \ttt\leq k(\om^*)\beta. $$
Consequently, in accordance with (2.5), for $\om^*\in\cW^*_\td(x)$ with $x\in\Lam_0$ and
$\ttt\in [0,k(\om^*)\beta ]$ the point ${\wt\om^*}(\ttt )= \om^*(\ttt )+s$. Therefore, the loops $\om^*$ from $\Om^*_0=\Om^*_\Lam\cap\cW^*_\td(\Lam_0)$
are shifted intact by the amount $s$ under the map (2.4). Consequently, the integral
energy $h(\Om^*_0)$ is not changed under tuned shifts.
\medskip

(v) The set $\tS(s)(\cD\cap \cG_L)$ will have a $\mu$-measure close to that of $\tS (s)\cD$;
moreover, the probability $\mu (\tS(s)(\cD\cap \cG_L))$ will be written in the form
$$\begin{array}{c}\mu (\tS(\pm s)(\cD\cap \cG_L))=\diy\int_{\cW^*_\td(\bbR^2)}
\mu (\rd\bOm^*_{\Lamc})
{\mathbf 1}\Big(\bOm^*_{\Lamc}\in\cW_\td\left(\Lamc \right)\Big)\\
 \diy\times\int_{\cW^*_\td(\Lam )}
\rd\Om^*_{\Lam}\;{\mathbf 1}\Big(\Om^*_\Lam\vee\bOm^*_{\Lamc}\in \cG_L\cap\cD\Big)\;
\frac{z^{K(\Om^*_\Lam)}}{L(\Om^*_\Lam)}\\
\qquad\qquad\qquad\times J^\pm_L(\Om^*_\Lam\vee\bOm^*_{\Lamc})
\exp\,\big[-h(\tT^\pm_L(s)\Om^*_{\Lam }|\bOm^*_{\Lamc})\,\big]
\end{array}\eqno (2.6)$$
where function $J^\pm_L=J^\pm_{L,s}$ gives the Jacobian of transformation $\tT^{\,\pm}_L(s)$.
By virtue of properties above (cf. particularly (i) and (iv)), the impact of $tT_L$ upon the energy
$h(tT_L\Om^*_{\Lam }|\bOm^*_{\Lamc})$
will be felt through the LC $\Om^*_{\Lam\setminus\Lam_0}=
\Om^*_\Lam\cap\cW^*_\td(\Lam\setminus\Lam^0)$ only. (More precisely, through
a LC $\Om^*_{\Lam\setminus\Lam_{R(L)}}$ where $\Lam_{R(L)}=[-R(L),R(L)]^{\times 2}$
and $R(L)\nearrow\infty$ with $L$. See Eqn (3.2) below.) Essentially, the same remains true about
the Jacobian $J_L(\Om^*_\Lam\vee\bOm^*_{\Lamc})$.
\medskip

(vi) In fact, a detailed analysis shows that second-order incremental expressions
$$\Big[J^+_L(\Om^*_\Lam\vee\bOm^*_{\Lamc})
J^-_L(\Om^*_\Lam\vee\bOm^*_{\Lamc})\Big]^{1/2}\eqno (2.7)$$
and
$$\exp\;\Big[
h(\tT^+_L(s)\Om^*_{\Lam }|\bOm^*_{\Lamc})
+h(\tT^-_L(s)\Om^*_{\Lam }|\bOm^*_{\Lamc})-
2h(\Om^*_{\Lam }|\bOm^*_{\Lamc})\Big]\eqno (2.8)$$
are close to $1$. It turns out that this fact suffices for the assertion of Theorem 2.1.

\medskip

Formally, Theorem 2.1 is derived from

\medskip

\medskip

{\bf Theorem 2.2.} {\sl For any $\delta >0$ there exists $L^*_0=L^*_0(\delta )>0$ such that
for $L\geq L^*_0$
$$\begin{array}{l}\diy{\rm{(A)}}\qquad\qquad\qquad\quad \mu (\cG_L)=\int_{\cW^*_\td(\bbR^2)}\mu (\rd\Om^*_{\Lamc})
\;{\mathbf 1}\Big(\Om^*_{\Lamc}\in\cW^*_\td(\Lamc )\Big)\\
\qquad\diy\times
\int_{\cW^*_\td(\Lam )}\rd\Om^*_\Lam\;{\mathbf 1}\Big(\Om^*_\Lam\vee\Om^*_{\Lamc}
\in \cG_L\Big)\\
\qquad\qquad\qquad\qquad\diy\times\;\frac{z^{K(\Om^*_\Lam)}}{L(\Om^*_\Lam)}
\;\exp\,\big[-h(\Om^*_{\Lam }|\bOm^*_{\Lamc})\,\big]\geq 1-\delta .
\end{array}\eqno (2.9)$$

{\rm{(B)}} The probabilities $\mu (\tS(\pm s)(\cD\cap \cG_L))$ are represented in the form (2.6)
with the following properties:
$\forall$ $\Om^*_\Lam\in\cW^*_\td(\Lam )$,
$\bOm^*_{\Lamc}\in\cW^*_\td(\Lamc )$ with $\Om^*_\Lam\vee
\bOm^*_{\Lamc}\in \cG_L$;

{\rm{(Ca)}} $\Big[J^+_L(\Om^*_\Lam\vee\bOm^*_{\Lamc})
J^-_L(\Om^*_\Lam\vee\bOm^*_{\Lamc})\Big]^{1/2}\geq 1-\delta$;
\medskip

{\rm{(Cb)}} $ h(\tT^+_L(s)\Om^*_{\Lam }|\bOm^*_{\Lamc})
+h(\tT^-_L(s)\Om^*_{\Lam }|\bOm^*_{\Lamc})-
2h(\Om^*_{\Lam }|\bOm^*_{\Lamc})\leq \delta$.}

\medskip

\medskip

The proof of Theorem 2.2 is carried on in the next sections.

\medskip

\medskip

{\bf Remark.} It is the pair of inequalities (Ca), (Cb) (together with the definition of
the `good' set $\cG_L$) where one crucially uses the fact that the physical dimension of the
system equals $2$.

\medskip

\medskip

We now show how to deduce the statement of Theorem 2.1 from that of Theorem 2.2.
Owing to Theorem 2.2 (A), (B), we can write:
$$\begin{array}{l}
\diy\hbox{the LHS of (2.1)}+3\delta\\
\qquad\diy\geq \mu (\tS(s)(\cD\cap \cG_L))+\mu (\tS(-s)(\cD\cap \cG_L))-2\mu (\cD\cap \cG_L)\\
\qquad =\diy\int_{\cW^*_\td(\bbR^2)}
\mu (\rd\bOm^*_{\Lamc})
{\mathbf 1}\Big(\bOm^*_{\Lamc}\in W_\td\left(\Lamc \right)\Big)\\
 \qquad\qquad\diy\times\int_{\cW^*_\td(\Lam )}
\rd\Om^*_{\Lam}\;{\mathbf 1}\Big(\Om^*_\Lam\vee\bOm^*_{\Lamc}\in \cG_L\cap\cD\Big)\;
\frac{z^{K(\Om^*_\Lam)}}{L(\Om^*_\Lam)}\\
\qquad\times\Big\{J^+_L(\Om^*_\Lam\vee\bOm^*_{\Lamc})
\exp\,\big[-h(\tT^+_L\Om^*_{\Lam }|\bOm^*_{\Lamc})\,\big]\\
\diy +J^-_L(\Om^*_\Lam\vee\bOm^*_{\Lamc})
\exp\,\big[-h(\tT^-_L\Om^*_{\Lam }|\bOm^*_{\Lamc})\,\big] -2\exp\,\big[-h(\Om^*_{\Lam }|\bOm^*_{\Lamc})\,
\big]\Big\}.\end{array}\eqno (2.10)$$
Next, by the AM/GM inequality, the RHS of (2.10) is no less than
$$\begin{array}{c}
\diy 2\int_{\cW^*_\td(\bbR^2)}
\mu (\rd\bOm^*_{\Lamc})
{\mathbf 1}\Big(\bOm^*_{\Lamc}\in W_\td\left(\Lamc \right)\Big)\int_{\cW^*_\td(\Lam )}
\rd\Om^*_{\Lam}\;{\mathbf 1}\Big(\Om^*_\Lam\vee\bOm^*_{\Lamc}\in \cG_L\cap\cD\Big)\\
\diy\times\frac{z^{K(\Om^*_\Lam)}}{L(\Om^*_\Lam)}\;\Big(\Big[J^+_L(\Om^*_\Lam\vee\bOm^*_{\Lamc})
J^-_L(\Om^*_\Lam\vee\bOm^*_{\Lamc})\Big]^{1/2}\qquad\qquad{}\\
\diy\times\exp\Big\{-\big[h(\tT^+_L\Om^*_{\Lam }|\bOm^*_{\Lamc})
+h(\tT^-_L\Om^*_{\Lam }|\bOm^*_{\Lamc})\,\big]\big/2\Big\}
 -\exp\,\big[-h(\Om^*_{\Lam }|\bOm^*_{\Lamc})\,
\big]\Big).\end{array}\eqno (2.11)$$

Now, by virtue of Theorem 2.2 (A)--(C), the RHS of (2.11) is greater than or equal to
$$\begin{array}{l}
\diy 2[(1-\delta)e^{-\delta/2}-1]\int_{\cW^*_\td(\bbR^2)}
\mu (\rd\bOm^*_{\Lamc})
{\mathbf 1}\Big(\bOm^*_{\Lamc}\in W_\td\left(\Lamc \right)\Big)\\
 \qquad\qquad\diy\times\int_{\cW^*_\td(\Lam )}
\rd\Om^*_{\Lam}\;{\mathbf 1}\Big(\Om^*_\Lam\vee\bOm^*_{\Lamc}\in \cG_L\cap\cD\Big)\\
\qquad\qquad\qquad\diy\times\frac{z^{K(\Om^*_\Lam)}}{L(\Om^*_\Lam)}
\exp\,\big[-h(\Om^*_{\Lam }|\bOm^*_{\Lamc})\,
\big]\\
\qquad \diy =2[(1-\delta)e^{-\delta/2}-1]\mu (\cG_L\cap\cD)\\ \;\\
\qquad\qquad \geq 2[(1-\delta)e^{-\delta/2}-1](1-\delta )
.\end{array}\eqno (2.12)$$
Since $\delta$ can be made arbitrarily small, we obtain the inequality (2.2).
\vskip 1 truecm

\centerline{\large{\bf 3. Definition of transformations $\tT^\pm_L$}}
\vskip 1 truecm

As was said earlier, the maps $\bOm^*\mapsto
\tT^\pm_L\bOm^*_\Lam\vee\bOm^*_{\Lamc}$ are
determined by transforming the $\ttt$-sections $\{\tT^\pm_L\bOm^*_\Lam\}(\ttt )$ of the
LC $\bOm^*_\Lam$, for each $\ttt\in [0,\beta ]$. Denoting by $T^\pm_L=T^\pm_L(\pm s)$
the map acting on CCs from $\cC_\td(\Lam )$, we can write:
$$\begin{array}{cl}\{\tT^\pm_L\bOm^*_\Lam\vee\bOm^*_{\Lamc}\}(\ttt )&=
\{\tT^\pm_L\bOm^*_\Lam\}(\ttt )\vee\bOm^*_{\Lamc}(\ttt )\\
\;&\qquad =\big(T^\pm_L[\{\bOm^*_\Lam\}(\ttt )]\big)\vee\bOm^*_{\Lamc}(\ttt ).\end{array}\eqno (3.1)$$
Like before, we would like to stress that the way the $\ttt$-section $\{\bOm^*_\Lam\}(\ttt )$
is transformed depends on $\{\bOm^*_{\Lamc}\}(\ttt)$, although $\{\bOm^*_{\Lamc}\}(\ttt )$ itself is not moving
when $\bOm^*\in\cG_L$.

More precisely, set:
$$R(L)=\big(\log\,\log\;L\big)^{3/4},\;\;\Lam_{R(L)}=[-R(L),R(L)]^{\times 2},\eqno (3.2)$$
and introduce yet another intermediate square
$$\oLam =[-\oL,\oL\,]^{\times 2}\;\hbox{ where }\;\oL=L-L^{3/4}.\eqno (3.3)$$
We will assume that the  quadruple of squares $\Lam_0$, $\Lam_{R(L)}$, $\oLam$ and $\Lam$ satisfies
$$\Lam_0\subset\Lam_{R(L)}\subset\oLam\subset\Lam .$$
The transformed CC $T^\pm_L[\{\bOm^*_\Lam\}(\ttt )]\in\cC_\td(\Lam )$
is formed by points ${\wt\om}^{*\pm}_L(l\beta +\ttt)$ obtained, as a result of shifts in the (positive) horizontal
direction, from the points $\om^*(l\beta +\ttt)$ where $\ttt\in [0,\beta ]$, $l=0,\ldots ,k(\om^*)-1$ and
$\om^*\in\bOm^*_\Lam$:
$${\wt\om}^{*\pm}_L(l\beta +\ttt)=\om^*(l\beta +\ttt) \pm {\tp}_L(\om^*(l\beta +\ttt))s.
\eqno (3.4)$$
Here the scalar value ${\tp}_L(\om^*(l\beta +\ttt))\geq 0$
depends on CCs
$\{\bOm^*_\Lam\}(\ttt )$ and $\{\bOm^*_{\Lamc}\}(\ttt)$ and are constructed recursively; cf. \cite{R2}. When
$\om^*(l\beta +\ttt )\in\Lam\setminus\oLam$, we have that
$${\tp}_L(\om^*(l\beta +\ttt))=0\;\hbox{ and }\;{\wt\om}^{*\pm}_L(l\beta +\ttt)
=\om^*(l\beta +\ttt).$$
In other words, a loop $\om^*\in\bOm^*$
is affected only at points $\om^*(\ttt )$ lying in $\oLam$.

In the course of construction of values ${\tp}_L(\om^*(l\beta +\ttt))$, we employ the function
$u\in [0,\infty )\mapsto \tau_L(u )$ determined  as follows:
$$\tau_L(u)=\begin{cases}1,&\quad 0\leq u\leq R(L),\\
1-\diy\frac{Q(u -R(L))}{Q(L-R(L))},&\quad R(L)\leq u\leq\oL,\\
0,&\quad u\geq\oL,\end{cases}\eqno (3.5)$$
where, in turn,\footnote{Function $\tau_L$ was introduced in \cite{FP} and \cite{P} and
has been repeatedly used in the literature.}
$$Q(u)=\int_0^uq(v)\rd v,\;\hbox{ with }\;q(v)=\frac{1}{1\vee v|\log v|}.\eqno (3.6)$$

The values ${\tp}(\om^*(l\beta +\ttt))={\tp}_L(\om^*(l\beta +\ttt))$ are related to
results of a series of minimizations, over points $\om^*(l\beta +\ttt)\in\{\bOm^*\}(\ttt )\cap\oLam$, of
subsequently introduced functions ${\wt t}\,^{\,(j)}(\;\cdot\;;\ttt )={\wt t}\,^{(j)}_L(\;\cdot\;;\ttt )$.
Here $j$ runs from $0$ to $\sharp\,\left(\{\bOm^*\}(\ttt )\cap\oLam\right)$ and the functions are
$$\begin{array}{l}
\diy y\in\bbR^2\mapsto {\wt t}\,^{(j)}(y;\ttt )\in [0,1],\\ \;\\
\qquad\qquad\diy 0\leq j\leq \sum\limits_{\om^*\in\bOm^*}
\sum\limits_{0\leq l <k(\om^*)}{\mathbf 1}(\om^*(l\beta +\ttt )\in\Lam).
\end{array}\eqno (3.7)$$
The value $j=0$ marks an initial function $t^{(0)}(\;\cdot\;;\ttt )$,
and the values $j\geq 1$ provide an ordering for points  $\om^*(l\beta +\ttt)$ in the CC
$\{\bOm^*\}(\ttt )\cap\oLam$. Let us stress that the functions ${\wt t}^{(j)}_L(\;\cdot\;;\ttt )$
involve (generally speaking) the whole $\ttt$-section $\{\bOm^*\}(\ttt )$.

The initial function in the series, ${\wt t}^{(0)}_L(\;\cdot\;;\ttt )$, does not depend on $\ttt\in [0,\beta ]$
and is related to function $\tau =\tau_L$ from (3.5):
$${\wt t}^{(0)}(y;\ttt ):= \tau\left(|y|_{\rmm}\right). \eqno (3.8)$$
Here and below, $|\,\cdot\,|_{\rmm}$ stands for the max-norm: $|y|_{\rmm}=\max\,\big[\big|\ty^{(1)}\big|,
\big|\ty^{(2)}\big|\big]$, for $y=\big(\ty^{(1)},\ty^{(2)}\big)$.
\medskip

The definition of the next function, ${\wt t}^{(1)}_L(\;\cdot\;;\ttt )$,
involves a (multiple) minimum of auxiliary functions $m_{x,0}$, over the points  $x=\om^*(l\beta +\ttt )$
from the CC $\{\bOm^*\}(\ttt )\cap\oLamc$:
$${\wt t}^{(1)}(y;\ttt )={\wt t}^{(0)}(y;\ttt )\wedge{\wt m}^{(0)}(y;\ttt )\eqno (3.9)$$
where
$${\wt m}^{(0)}(y;\ttt )={\wt m}^{(0)}_L(y;\ttt )=\operatornamewithlimits{\bigwedge}\limits_{\om^*(l\beta +\ttt)\in
{\diy\{\bOm^*\}}(\ttt )\cap\oLamc}m_{\om^*(l\beta +\ttt ),0}\,(y)
.\eqno (3.10)$$
Here and below, following \cite{P}, \cite{FP}, \cite{R2}, the family of auxiliary functions
$y\in\bbR^2\mapsto m_{x,\tu}(y)$ is used,
with values in $[0,1)\cup\{+\infty\}$, where $x\in\bbR^2$, $\tu\in [0,1)$.
These functions are introduced as follows:
$$m_{x,\tu}(y):=\begin{cases}\tu, &\tth_{x,\tu}\tc_f>1/2,\\
\tu+\tth_{x,\tu}f(x-y)&\;\\
\qquad+\infty\cdot{\mathbf 1}(f(x-y)=1), &\tth_{x,\ttr}\tc_f\leq1/2.
\end{cases}\eqno (3.11)$$
In turn, $f=f_\eps$ is a chosen C$^1$-function $\bbR^2\to [0,1]$, with
$$f(v)=0\;\hbox{ when }\;|v|<a\;\hbox{ and }\;f(v)=1\;\hbox{
when }\;|v|>a+2\eps ,$$ and
$$\tc_f=\max\,\Big[|\nabla f(v)|,\;v\in\bbR^2\Big]\,.\eqno (3.12)$$
The value $\eps$ is selected for given $z$
and $\beta$ satisfying (1.4) and should be small enough, guaranteeing smallness of
quantities introduced below. Finally,
$$\tth_{x,\tu}:=|\tau (|x|_{\rmm}-\eps -a/2)-\tu\,|\eqno (3.13)$$
is another auxiliary parameter.

Pictorially speaking, the function $y\in\bbR^2\mapsto {\wt m}^{(0)}(y;\ttt )$ indicates
by how much a particle (i.e., a circle of diameter $\td $) placed at the reference point $y$ could be moved (under adopted
arrangements) in presence of hard-core particles
placed at points $\om^*(l\beta +\ttt )\in\{\bOm^*\}(\ttt )\cap\oLamc$. Consequently,
${\wt t}^{(1)}_L(y;\ttt )$ indicates how much a movement by quantity ${\wt t}^{(0)}_L(y;\ttt )$
should be reduced in presence of hard-core particles
at $\om^*(l\beta +\ttt )\in\{\bOm^*\}(\ttt )\cap\oLamc$. We then look for the minimum
of ${\wt t}^{(1)}_L(\;\cdot\;;\ttt )$ over the CC  $\{\bOm^*\}(\ttt )\cap\oLam$
and set:
$$\begin{array}{c}\tp^1=\tp^1_L=\min\,\Big[{\wt t}^{(1)}_L(y;\ttt ):\;y\in\{\bOm^*\}(\ttt )\cap\oLam\Big],\\
P^1=P^1_L=\hbox{arg min}\,\Big[{\wt t}^{(1)}_L(y;\ttt ):\;y\in\{\bOm^*\}(\ttt )\cap\oLam\Big].\end{array}
\eqno (3.14)$$
If the minimum is attained at more than one point in $\{\bOm^*\}(\ttt )\cap\oLam$, we
list all these points: $P^1$, $\ldots$, $P^{\kappa_1}$ (in any order).
The value $\tp^1$ is assigned to each of those points as ${\tp}(P^j)$:
$$\begin{array}{c}{\tp}(\om^*(l\beta +\ttt ))=\tp^1,\;\hbox{ if }\;
{\om^*}\in\bOm^*,0\leq l<k(\om^*),\\
\qquad\qquad\qquad\quad\om^*(l\beta +\ttt )\in\oLam\;
\hbox{ and }\;{\wt t}^{(1)}(\om^*(l\beta +\ttt );\ttt )=\tp^1.\end{array}\eqno (3.15)$$

The value $\tp^1$ and the position $P^1$ (or the positions  $P^1$, $P^2$, $\ldots$, $P^{\kappa_1}$)
are taken into account  in the definition of the next function $y\in\bbR^2\mapsto{\wt t}^{(2)}(y;\ttt )$:
$$\begin{array}{cl}{\wt t}^{(2)}(y;\ttt )
&={\wt t}^{(1)}(y;\ttt )\wedge m_{P^1,\tp^1\times\ts}(y)\ldots\wedge m_{P^{\kappa_1},\tp^1\times\ts}(y)\\
\;&\;\\
\;&=
{\wt t}^{(0)}(y;\ttt )\wedge{\wt m}^{(1)}(y;\ttt ).\end{array}\eqno (3.16)$$
Here ${\wt m}^{(1)}(y;\ttt )={\wt m}^{(1)}_L(y;\ttt )$ is given by
$${\wt m}^{(1)}(y;\ttt )={\wt m}^{(0)}(y;\ttt )\wedge\left(
\operatornamewithlimits{\bigwedge}\limits_{\om^*(l\beta +\ttt)\in
{\diy\{\bOm^*\}^1}(\ttt )\cap\oLam}m_{\om^*(l\beta +\ttt ),\tp^1\times\ts}\,(y)\right)
\eqno (3.17)$$
and
$$\{\bOm^*\}^1(\ttt )=
\Big\{\om^*(l\beta +\ttt )\in\{\bOm^*\}(\ttt ):\;
{\wt t}^{(1)}(y;\ttt )=\tp^1\Big\}\eqno (3.18)$$
yielding that
$$\{\bOm^*\}^1(\ttt )\cap\oLam =\{P^1,\ldots ,P^{\kappa_s}\}.$$
(Recall, the initial shift-vector is $s=(\ts ,0)$ where $\ts\in [0,1/2)$.)

Pictorially, as before, the function $y\in\bbR^2\mapsto {\wt m}^{(1)}(y;\ttt )$ indicates
by how much a particle at point $y$ could be moved when we take into account the particles
placed at points $\om^*(l\beta +\ttt )\in\{\bOm^*\}(\ttt )\cap\oLamc$ (which do not move)
and the particles placed at points  $\om^*(l\beta +\ttt )\in\{\bOm^*\}^1(\ttt )\cap\oLam$
(which are moved by $\tp^1$). Consequently,
${\wt t}^{(2)}(y;\ttt )$ indicates how much a movement by quantity ${\wt t}^{(0)}(y;\ttt )$
should be reduced in presence of hard-core particles
at points $\om^*(l\beta +\ttt )\in\{\bOm^*\}(\ttt )\cap\oLamc$ and
$\om^*(l\beta +\ttt )\in\{\bOm^*\}^1(\ttt )\cap\oLam$.

Next, we minimise the function ${\wt t}^{(2)}(\;\cdot\;;\ttt )$ over the $\ttt$-section\\
$\left(\{\bOm^*\}(\ttt )\setminus\{\bOm\}^1(\ttt )\right)\cap\oLam$
and, like before, set:
$$\begin{array}{c}\tp^2=\min\,\Big[{\wt t}^{(2)}_L(y;\ttt ):\;y\in\left(\{\bOm^*\}(\ttt )\setminus\{\bOm\}^1(\ttt )\right)
\cap\oLam\Big],\\
P^{\kappa_1+1}=\hbox{arg min}\,\Big[{\wt t}^{(2)}(y;\ttt ):\;y\in\left(\{\bOm^*\}(\ttt )\setminus\{\bOm\}^1(\ttt )\right)\cap\oLam\Big].\end{array}
\eqno (3.19)$$
Again, if the minimum is shared by more than one point in\\
$\left(\{\bOm^*\}(\ttt )\setminus\{\bOm\}^1(\ttt )\right)\cap\oLam$, we
list all these points: $P^{\kappa_1+1}$,  $\ldots$, $P^{\kappa_1+\kappa_2}$ (in any order).
As earlier, the value $\tp^2$ is assigned to each of those points as ${\tp}(P^j)$:
$$\begin{array}{c}{\tp}(\om^*(l\beta +\ttt ))=\tp^2,\;\hbox{ if }\;
{\om^*}\in\bOm^*,0\leq l<k(\om^*),\\
\qquad\qquad\qquad\quad\om^*(l\beta +\ttt )\in\oLam\;
\hbox{ and }\;{\wt t}^{(2)}(\om^*(l\beta +\ttt );\ttt )=\tp^2.\end{array}$$

And so on: this procedure is iterated until we exhaust all points in $\{\bOm^*\}(\ttt )\cap\oLam$.
(Recall, their number and their positions vary with $\ttt\in [0,\beta ]$.) At the end, we obtain a
resulting function ${\wt t}={\wt t}_L (\;\cdot\;;\ttt )$:
$$y\in\bbR^2\mapsto{\wt t}(y)\;\hbox{ where }\;{\wt t}(y)={\wt t}^{(0)}(y)\wedge{\wt m}(y)
\eqno (3.20)$$
where
$${\wt m}(y)={\wt m}_L(y;\ttt )=\operatornamewithlimits{\bigwedge}\limits_{\om^*(l\beta +\ttt)\in
{\diy\{\bOm^*\}}(\ttt )}m_{\om^*(l\beta +\ttt ),{\tp}(\om^*(l\beta +\ttt ))\times\ts}\,(y)
.\eqno (3.21)$$
Here we set:
$${\tp}(\om^*(l\beta +\ttt ))=0\;\hbox{ when }\;\om^*(l\beta +\ttt )\in\oLamc .$$
Observe that
$${\wt t}_L(y;\ttt )=0\;\hbox{for}\;y\in\oLamc\;\hbox{ and }\;{\wt t}_L(y;\ttt )=1\;\hbox{for}\;y\in\Lam_{R(L)}.
\eqno (3.22)$$

\medskip

The Jacobian $J^\pm_L(\Om^*_\Lam\vee\bOm^*_{\Lamc})$ of the transform $\tT^\pm _L$
turns out to be of the form:
$$\begin{array}{l}
\diy J^\pm_L(\Om^*_\Lam\vee\bOm^*_{\Lamc})=\exp\,\Bigg[\int_0^\beta \rd\ttt\,
\sum\limits_{\om^*\in{\diy\Om^*_\Lam\vee\bOm^*_{\Lamc}}}\\
\qquad\qquad\qquad\diy\times\sum\limits_{0\leq l<k^*(\om )}\ln\,\Bigg(1\diy\pm\ts\times
\big(\partial^1{\wt t}_L\,\big)(\om^*(l\beta +\ttt );\ttt)\Bigg)\Bigg]\end{array}\eqno (3.23)$$
where $(\partial^1{\wt t}_L\,)(y)$ stands for the partial derivative
$\diy\frac{\partial {\wt t}_L}{\partial{\ty}^1}(y;\ttt )$, $y=(\ty^1,\ty^2)$.
(The fact that the functions ${\wt t}_L$ are non-differentiable on sets of positive co-dimension
is not an obstacle here because of involvement of  Wiener's integration.)
The crucial quantity $\Big[J^+_L(\Om^*_\Lam\vee\bOm^*_{\Lamc})
J^-_L(\Om^*_\Lam\vee\bOm^*_{\Lamc})\Big]^{1/2}$ in Eqn (2.11) becomes
$$\begin{array}{l}\diy
\Big[J^+_L(\Om^*_\Lam\vee\bOm^*_{\Lamc})
J^-_L(\Om^*_\Lam\vee\bOm^*_{\Lamc})\Big]^{1/2}=\exp\,\Bigg(\int_0^\beta \rd\ttt\\
\qquad\diy\times\sum\limits_{\om^*\in{\diy\Om^*_\Lam\vee\bOm^*_{\Lamc}}}
\sum\limits_{0\leq l<k^*(\om )}\ln\,\Bigg\{1-
\Big[\ts^2\big(\partial^1{\wt t}_L\big)(\om^*(l\beta +\ttt );\ttt)\Big]^2\Bigg\}\Bigg).\end{array}\eqno (3.24)$$
We see that the quantity (3.24) is close to $1$ when we are able to check
that the sum
$$\sum\limits_{\om^*\in{\diy\Om^*_\Lam\vee\bOm^*_{\Lamc}}}
\sum\limits_{0\leq l<k^*(\om )}\int_0^\beta \rd\ttt
\Big[\big(\partial^1{\wt t}_L\big)(\om^*(l\beta +\ttt );\ttt)\Big]^2\eqno (3.25)$$
is close to $0$.

\medskip

We conclude this section with a straightforward assertion justifying the definition
(3.3) that introduces the intermediate square $\oLam$.

\medskip

\medskip

{\bf Lemma 3.1.} {\sl Consider the events
$$\cL^{(1)}_L=\{\bOm^*\in\cW^*_\td(\bbR^2 ):\;\alpha_{\bbR^2\setminus\oLam}(\om^*)=1\;
\;\forall\;\om^*\in\bOm^*\;\hbox{ with }\;x(\om^*)\in\Lamc\}\eqno (3.25.1)$$
and
$$\cL^{(2)}_L=\{\bOm^*\in\cW^*_\td(\bbR^2 ):\;\alpha_{\Lam_{R(L)}}(\om^*)=1\;
\;\forall\;\om^*\in\bOm^*\;\hbox{ with }\;x(\om^*)\in\Lam_0\}\eqno (3.25.2)$$

In other words, {\rm{(a)}} for $\bOm^*\in\cL^{(1)}_L$, every loop $\om^*$ from $\bOm^*$ which starts
at a point $x(\om^*)$ outside square $\Lam$ does not reach square $\oLam$, while {\rm{(b)}} for
$\bOm^*\in\cL^{(2)}_L$, every loop $\om^*_0$ from $\bOm^*_{\Lam_0}$ (which starts
in $\Lam_0$) does not leave square $\Lam_{R(L)}$.
Then, under condition {\rm{(1.4)}},
$$\lim_{L\to\infty}\mu_L(\cL^{(1)}_L)=\lim_{L\to\infty}\mu_L(\cL^{(2)}_L)=1,\eqno (3.26)$$
$\forall$ $\mu\in\fK (z,\beta )$.}

\medskip

\medskip

{\it Proof of Lemma} 3.1. Both relations are proved in a similar way, so we discuss
in detail one of them, say $\lim\limits_{L\to\infty}\mu_L(\cL^{(1)}_L)=1$. At first we write
$$\begin{array}{l}
\mu (\cW^*_\td\setminus\cL^{(1)}_L)\\
\quad =\mu (\hbox{$\exists$ at least one loop $\om^*$ with
$x(\om^*)\in\Lamc$ reaching $\oLam$})\\
\quad\diy \leq \int\mu (\rd\bOm^*)\sum\limits_{\om^*\in\bOm^*_{\Lamc}}{\mathbf 1}\Big(
\om^*(\ttt )\in\oLam\;\hbox{for some }\;\ttt\in [0,k(\om^*)\beta]\Big).\end{array}$$
By virtue of the Campbell theorem, the last integral equals
$$\int\rd\om^*\rho (\om^*)\\
{\mathbf 1}\Big(x(\om^*)\in\Lamc\;\hbox{but}\;
\om^*(\ttt )\in\oLam\;\hbox{for some }\;\ttt\in [0,k(\om^*)\beta]\Big).$$
By the Ruelle bound (cf. Eqns (2.3.18.{\bfI})--(2.3.20.{\bfI})) this does not exceed
$$\begin{array}{l}\diy
\quad\diy \int_{\Lamc}\rd x\int_{\cW^*(x)}\bbP_x(\rd\om^*)\frac{\orho^{k(\om^*)}}{k(\om^*)}
{\mathbf 1}\Big(
\om^*(\ttt )\in\oLam\;\hbox{for some }\;\ttt\in [0,k(\om^*)\beta]\Big).
\end{array}$$

Next, we observe that the loop $\om^*$ with the endpoint $x=(\tx^1,\tx^2)\in\Lamc$
(i.e., with $\max\,|\tx^j|_{\rmm}\geq L$) can reach $\oLam$ only if
at least one of its one-dimensional components (i.e., a scalar Brownian bridge
with the endpoint $\tx^j$, $j=1$ or $2$) deviates from its origin by at least $(|\tx^j|-L)+L^{3/4}$.
Therefore, the last displayed expression is upper-bounded by
$$\begin{array}{l}
\diy 2\times 2\sum\limits_{k\geq 1}\frac{\orho^k}{k\sqrt{2\pi\beta k}}\int_{L^{3/4}}^\infty\rd\tx
\exp\,\big[-4\tx^2/(2k\beta)\big]\\
\qquad\diy\leq\sum\limits_{k\geq 1}\frac{4\orho^k}{{\sqrt{2\pi}}\,k}\,\frac{\exp\,[-L^{3/2}/(2k\beta)]}{L^{3/4}/\sqrt{k\beta}+
\sqrt{L^{3/2}/(k\beta )\,+4/\pi}}\,.
\end{array}\eqno (3.27)$$
Here we have used an estimate for the (scalar) Brownian bridge $\tB (\ttt )$ with endpoints $\t0$
and $\ty\geq 0$: $\forall$ $\tA>\ty$
$$\otP^{\beta k}_{0,\ty}\Big\{\sup\;[\tB (\ttt ):\;0\leq \ttt\leq \beta k]\geq\tA\Big\}=
\frac{1}{\sqrt{2\pi\beta k}}e^{-(2\tA-y)^2/(2\beta k)}\eqno (3.28.1)$$
plus well-known estimates for the tail of the normal distribution (Mills ratio bounds): $\forall$ $\tA\in (0,\infty )$,
$$\begin{array}{c}
\diy\frac{e^{-{\tA}^2/2}}{{\tA}+\sqrt{{\tA}^2+2}}\leq\int_{\tA}^{\infty}
e^{-\ttt^2/2}\rd\tt\ttt\leq \frac{e^{-{\tA}^2/2}}{{\tA}+\sqrt{{\tA}^2+4/\pi}}.\end{array}
\eqno (3.28.2)$$

It is not hard to see that the RHS of (3.27) tends to $0$ as $L\to\infty$. This completes the proof
of Lemma 3.1.

In what follows we will assume that an LC $\bOm^*$ lies in $\cL_L$. Together with
(3.22) this will imply that the loops $\om^*\in\bOm^*$ with $x(\om^*)\in\Lamc$ remains unaffected
by transformations $\tT^\pm(s)$.
\medskip

\medskip

\vskip 1 truecm

\centerline{\large{\bf 4. Estimates for the Jacobians}}
\vskip 1 truecm

To guarantee assertions (A) and (Ca) of Theorem 2.2 we need to secure that
the good set $\cG_L$ carries a large measure and contains only those
LCs $\bOm^*\in\cW^*(\bbR^2)$ for which the expression
$J^+_L(\Om^*_\Lam\vee\bOm^*_{\Lamc})J^-_L(\Om^*_\Lam\vee\bOm^*_{\Lamc})$ can be
appropriately controlled. To this end, consider a random variable $\Sigma^{\rJ}(\bOm^*)=
\Sigma^{\rJ}_L(\bOm^*)$ given by the RHS of (3.24):
$$\begin{array}{l}\diy\Sigma^{\rJ}(\bOm^*):
\bOm^*\mapsto \int_0^\beta \rd\ttt\sum\limits_{x\in\{\bOm^*\}(\ttt )}
\Big[\big(\partial^1{\wt t}_L\big)(x;\ttt )\Big]^2\\
\qquad\qquad\qquad\diy =\;
\sum\limits_{\om^*\in{\diy\bOm^*}}\int_0^\beta \rd\ttt\,\sum\limits_{0\leq l<k^*(\om )}\,
\Big[\big(\partial^1{\wt t}_L\big)(\om^*(l\beta +\ttt );\ttt)\Big]^2
.\end{array}\eqno (4.1)$$
The formal definition of the set $\cG_L$ will require that the quantity $\Sigma^{\rJ}(\bOm^*)$
is small (more precisely that some majorants for $\Sigma^{\rJ}(\bOm^*)$ are small); see below.
Formally, the property that $J^+_L(\Om^*_\Lam\vee\bOm^*_{\Lamc})J^-_L(\Om^*_\Lam\vee\bOm^*_{\Lamc})$
is close to $1$  follows from

\medskip

\medskip

{\bf Lemma 4.1.} {\sl If $\eps >0$ is chosen small enough then
the mean-value of $\Sigma^{\rJ}(\bOm^*)$ vanishes as $L\to\infty$:
$$
\lim\limits_{L\to\infty}\int\;\mu (\rd\bOm^*)\Sigma^{\rm J}(\bOm^*)=0.
\eqno (4.2)$$}

\medskip

\medskip

{\it Proof of Lemma} 4.1. Let us start with technical definitions.
Given $\ttt\in [0,\beta ]$ and $x,x'\in\{\bOm^*\}(t)$, we write:
$$\left.\begin{array}{cl}x\leftrightarrow x'&\hbox{whenever $a<|x-x'|<a+\eps$}\\
\quad\hbox{and}&\;\\
 x{\buildrel\;\{\bOm^*\}(\ttt)\over{\longleftarrow
\longrightarrow}}x''&\hbox{when there exists  a collection of particles}\\
\;&\hbox{$x_0$, $\ldots$, $x_m\in\bOm^*$ such that point $x_0$ coincides}\\
\;&\hbox{with $x$, point
$x_m$ with $x''$ and $\forall$ $i=1,\ldots ,m$}\\
\;&\hbox{$x_{i-1}$ and $x_i$
satisfy $a<|x_i-x_{i-1}|<a+\eps$.}\end{array}\right\}\eqno (4.3)$$

Recall, the values
$z,\beta >0$ are such that the bound (1.4) is satisfied. Referring below
to a small $\eps >0$ we mean conditions like this:
$$\frac{1}{4}\pi [(a+2\epsilon)^2-a^2](1\vee\beta)\left(1\vee
\sum\limits_{k\geq 1}
\orho^kk\Big/(2\pi\beta )\right)<1.\eqno (4.4)$$

To assess the integral in (4.2), observe that one possibility for value ${\wt t}_L(\om^*(l\beta +\ttt ))$
is ${\wt t}^{(0)}_L(\om^*(l\beta +\ttt ))$; the opposite case is where
${\wt t}_L(\om^*(l\beta +\ttt ))$ equals  ${\wt m}_L(\om^*(l\beta +\ttt ))$. See Eqns (3.8), (3.20).
In the former case we have to deal with the derivative
$$\partial^1{\wt t}^{(0)}_L(\om^*(l\beta +\ttt);\ttt)=Z_L(|\om^*(l\beta +\ttt )|_{\rmm})
$$
where
$$Z_L(r)=\frac{\left[q\left(r-R(L)-\eps-a/2\right)\right]^2}{
\left[Q(\oL-R(L)-\eps-a/2)\right]^2}\,{\mathbf 1}(0\leq r\leq\oL).\eqno (4.5)$$
In the second case we obtain that
$${\wt t}_L(\om^*(l\beta +\ttt ))={\wt m}_L(\om^*(l\beta +\ttt )),$$
and we have to use the structure of function ${\wt m}_L(\om^*(l\beta +\ttt ))$
(related to multiple minimisation as defined in Eqn (3.21)) to assess its derivative.
Cf. Sect 6.7 in \cite{R2}.

All in all, to verify (4.2) it suffices to check that
$$\lim\limits_{L\to\infty}\int\;\mu (\rd\bOm^*)[\Sigma^{(1)}(\bOm^*)+\Sigma^{(2)}(\bOm^*)]
=0.\eqno (4.6)$$
Here variable $\Sigma^{(1)}=\Sigma^{(1)}_L$ is given by
$$\begin{array}{cl}\Sigma^{(1)}(\bOm^*)&\diy =\int_0^\beta\rd\ttt
\sum_{x\in\{\bOm^*\}(\ttt )}\tau_L(|x|_{\rmm})\\
\;&\diy =\sum_{\om^*\in \bOm^*}
\int_0^{k(\om^*)\beta}\tau_L(|\om^*(\ttt )|_{\rmm}) \rd\ttt\end{array}\eqno (4.7)$$
and corresponds to the first of the aforementioned possibilities (where we have ${\wt t}_L(\om^*(l\beta +\ttt ))
={\wt t}^{(0)}_L(\om^*(l\beta +\ttt ))$). The function $\tau_L$ 
Next, variable $\Sigma^{(2)}=\Sigma^{(2)}_L$ corresponds with the second possibility and has the form
$$\begin{array}{l}\diy \Sigma^{(2)}(\bOm^*)
=\int_0^\beta\rd\ttt
\sum\limits_{\substack{x,x',x''\in\{\bOm^*\}(\ttt )\\ x\neq x'}}
{\mathbf 1}\Big(x\leftrightarrow x'\Big)
{\mathbf 1}\Big(x{\buildrel\;\{\bOm^*\}(\ttt )\over{\longleftarrow\longrightarrow}}x''\Big)\\
\qquad\quad\diy\times{\mathbf 1}\Big(|x|_{\rmm}\leq |x''|_{\rmm}\Big)
\Big[\tau_L( |x|_{\rmm}-\eps -a/2)-\tau_L( |x''|_{\rmm})\Big]^2\\ \;\\
\qquad\qquad :=\Sigma^{(2,1)}(\bOm^*)+\Sigma^{(2,2)}(\bOm^*).
\end{array}\eqno (4.8)$$
The composition of the RHS is related to a `cluster' structure accompanying
the multiple minimization procedure in (3.21) which determines the value of
interest ${\wt m}_L(\om^*(l\beta +\ttt ))$.
Formally, as follows from the definition, behind
the indicator ${\mathbf 1}\Big(x{\buildrel\;\{\bOm^*\}(\ttt )\over{\longleftarrow\longrightarrow}}x''\Big)$
there is a `chain' of points from the $\ttt$-section $\{\bOm^*\}(\ttt )$ which joins the
`extreme' points $x$ and $x''$.
Cf. Sect. 8 in \cite{R2} (whose system notation is partially followed here).

Moreover, the partition $\Sigma^{(2)}(\bOm^*)=\Sigma^{(2,1)}(\bOm^*)+\Sigma^{(2,2)}(\bOm^*)$
reflects the fact that $x$ and $x''$, the two extreme points in the chain, can belong to the same
loop $\om^*$ or to two distinct loops, $\om^*$ and ${\om^*}''$. More precisely,
the summand $\Sigma^{(2,1)}=\Sigma^{(2,1)}_L$ is specified as the sum
$$\begin{array}{l}\diy
\sum\limits_{\om^*\in\bOm^*}\int_0^\beta\rd\ttt
\sum_{0\leq l,l''<k(\om^*)}{\mathbf 1}\left(\om^*(l\beta +\ttt )
{\buildrel\{\bOm^*\}(\ttt)\over{\longleftarrow\longrightarrow}}\om^*(l''\beta+\ttt )\right)\\
\quad\diy\times\bigg[{\mathbf 1}(\om^*(l\beta +\ttt )
\leftrightarrow\om^*(l''\beta+\ttt )){\mathbf 1}(l\neq l'') \\
\qquad\diy + \sum\limits_{{\om^*}'\neq{\om^*}''}\sum\limits_{0\leq l'<k({\om^*}')}{\mathbf 1}
(\om^*(l\beta +\ttt )\leftrightarrow{\om^*}'(l'\beta +\ttt ))\bigg]\\
\qquad\quad\diy\times{\mathbf 1}\Big(|\om^*(l\beta +\ttt )|_{\rmm}\leq |\om^*(l''\beta+\ttt )|_{\rmm}\Big)\\
\quad\diy\times\Big[\tau_L( |\om^*(l\beta +\ttt )|_{\rmm}-\eps -a/2)-\tau_L( |\om^*(l''\beta+\ttt )|_{\rmm})\Big]^2
\end{array}\eqno (4.9.1)$$
whereas the term $\Sigma^{(2,2)}=\Sigma^{(2,2)}_L$ equals the sum
$$\begin{array}{l}
\diy\sum\limits_{\om^*,{\om^*}''\in\bOm^*}\int_0^\beta\rd\ttt
\sum_{\substack{0\leq l<k(\om^*)\\ 0\leq l''<k({\om^*}'')}}{\mathbf 1}\left(\om^*(l\beta +\ttt )
{\buildrel\{\bOm^*\}(\ttt)\over{\longleftarrow\longrightarrow}}{\om^*}''(l''\beta+\ttt )\right)\\
\quad\diy\times\bigg[{\mathbf 1}(\om^*(l\beta +\ttt )
\leftrightarrow{\om^*}''(l''\beta+\ttt ))\\
\quad\diy + \sum\limits_{0\leq l'<k({\om^*}'')}{\mathbf 1}(l'\neq l''){\mathbf 1}
(\om^*(l\beta +\ttt )\leftrightarrow{\om^*}''(l'\beta +\ttt ))\\
\qquad\diy+ \sum\limits_{{\om^*}'\neq{\om^*}''}\sum\limits_{0\leq l'<k({\om^*}')}{\mathbf 1}
(\om^*(l\beta +\ttt )\leftrightarrow{\om^*}'(l'\beta +\ttt ))\bigg]\\
\qquad\quad\diy\times{\mathbf 1}\Big(|\om^*(l\beta +\ttt )|_{\rmm}\leq |{\om^*}''(l''\beta+\ttt )|_{\rmm}\Big)\\
\quad\diy\times\Big[\tau_L( |\om^*(l\beta +\ttt )|_{\rmm}-\eps -a/2)-\tau_L( |{\om^*}''(l''\beta+\ttt )|_{\rmm})\Big]^2.
\end{array}\eqno (4.9.2)$$

\medskip

Constants $C_j\in (0,\infty )$ appearing in the argument below vary with $\beta$ and $z$ (through
$\orho$) but  are independent of $L$.

\medskip

{\bf Proposition 4.1.} {\sl The mean value of $\Sigma^{(1)}$ is assessed as follows:
$$\int_{\cW^*_\td(\bbR^2)}\mu (\rd\bOm^*)\Sigma^{(1)}_L(\bOm^*)\leq C_0 \gamma (L)\eqno (4.9)$$
where $C_0\in (0,\infty )$ is a constant and the quantity $\gamma (L)$ is defined as
follows:}
$$\gamma (L):=\int_{\oLam}\frac{q(|x|_{\rmm}-R(L)-\eps -a/2 )^2}{Q(L-R(L)-\eps-a/2)^2}\rd x,
\;\hbox{ with }\;\diy\lim\limits_{L\to\infty}\gamma (L)=0.\eqno (4.10)$$

\medskip

\medskip

{\it Proof of Proposition} 4.1. To explain the bound (4.10),  we first write, by the Campbell theorem:
$$\begin{array}{l}
\diy\int_{\cW^*_\td(\bbR^2)}\mu (\rd\bOm^*)\Sigma^{(1)}_L(\bOm^*)\\
\qquad\diy =\int_{\cW^*}\rd\om^*\rho (\om^*)
\int_0^{k(\om^*)\beta}\tau_L(|\om^*(\ttt )|_{\rmm})\rd\ttt .\end{array}\eqno (4.11)$$

By the Ruelle bound, the RHS does not exceed
$$\int_{\cW^*}\rd\om^*\frac{\orho^{k(\om^*)}}{k(\om^*)}
\int_0^{k(\om^*)\beta}\tau_L(|\om^*(\ttt )|_{\rmm})\rd\ttt  .\eqno (4.12)$$
When $x(\om^*)\in\Lam_{R(L)}$, we estimate
$$Z_L(|\om^*(\ttt )|_{\rmm})\leq\diy\frac{1} {\left[Q(L-R(L)-\eps-a/2)\right]^2};
\eqno (4.13)$$
consequently, the corresponding contribution
$$\begin{array}{l}
\diy\int_{\cW^*}\rd\om^*{\mathbf 1}(x(\om^*)\in\Lam_{R(L)})
\frac{\orho^{k(\om^*)}}{k(\om^*)}
\int_0^{k(\om^*)\beta}\tau_L(|\om^*(\ttt )|_{\rmm})\rd\ttt  \end{array}$$
does not exceed
$$\begin{array}{l}\diy\frac{(2R(L))^2}{\left[Q(L-R(L)-\eps-a/2)\right]^2}
\sum\limits_{k\geq 1}
\frac{(k\beta)\orho^k}{(2\pi k\beta)k}\\
\qquad\diy = \frac{\big(\log\,\log\;L\big)^{3/2}}
{2\pi\left[Q(L-R(L)-\eps-a/2)\right]^2}
\sum\limits_{k\geq 1}
\frac{\orho^k}{k}\\
 \qquad\qquad\diy< \frac{\orho\big(\log\,\log\;L\big)^{3/2}}{2\pi(1-\orho )
 \left[Q(L-R(L)-\eps-a/2)\right]^2} .\end{array}\eqno (4.14)$$

This idea can be pushed further: we use estimate (4.14) whenever
loop $\om^*$ reaches $\Lam_{R(L)}$. For given $x\not\in\Lam_{R(L)}$ and $\om^*\in\cW^*(x)$
this can occur when either (i) $k(\om^*)$ is large -- say, $k(\om^*)>\big[|x|_{\rmm}-R(L)\big]\big/2$ -- or
when  (ii) the opposite inequality $k(\om^*)\leq \big[|x|_{\rmm} -R(L)\big]\big/2$ holds true but
the loop $\om^*$ deviates from $x$, in the max-distance, by at least $|x|_{\rmm} -R(L)$. Then
the corresponding part of expression (4.14)
$$\begin{array}{l}
\diy\int_{\cW^*}\rd\om^*{\mathbf 1}(x(\om^*)\not\in\Lam_{R(L)})\\
\qquad\times{\mathbf 1}(\om^*(\ttt )\in\Lam_{R(L)}\;
\hbox{ for some }\;t\in [0,k(\om^*)\beta ])\\
\qquad\qquad\qquad\qquad\quad\diy\times\frac{\orho^{k(\om^*)}}{k(\om^*)}
\int_0^{k(\om^*)\beta}\tau_L(|\om^*(\ttt )|_{\rmm})\rd\ttt  \end{array}$$
is upper-bounded by
$$\begin{array}{l}\diy\int_{\bbR^2} \rd x \bigg\{\sum\limits_{k\geq |x|_{\rmm}/2}
\frac{(k\beta)\orho^k}{(2\pi k\beta )k}+\sum\limits_{1\leq k\leq |x|_{\rmm}/2}
\frac{(k\beta)\orho^k}{k}\int_{\cW^{k\beta}(0)}\bbP^{k\beta}_0(\rd\om^*)\\
\qquad\quad\diy\times
{\mathbf 1}\Big(\max\,\big[|\om^*(\ttt )|_{\rmm}:\;0\leq \ttt\leq k\beta\big]>|x|_{\rmm}\Big)\bigg\}.
\end{array}\eqno (4.15)$$

The first sum in (4.15) is evaluated through a convergent geometric progression:
$$\begin{array}{c}\diy\sum\limits_{k\geq |x|_{\rmm}/2}
\frac{\orho^k}{2\pi k}\leq\frac{\orho^{|x|_{\rmm}/2}}{
2\pi (1-\orho )}\,,\end{array}$$
and its contribution into the integral $\diy\int_{\bbR^2} \rd x$ does not exceed a constant. To estimate
the second sum, one can use the inequalities (3.27.1,2).
This yields:
$$\begin{array}{l}
\diy\sum\limits_{1\leq k\leq |x|^{1/2}_{\rmm}}
\frac{(k\beta)\orho^k}{k}\int_{\cW^{k\beta}(0)}\bbP^{k\beta}_0(\rd\om^*)\\
\qquad\quad\diy\times
{\mathbf 1}\Big(\max\,\big[|\om^*(\ttt )|_{\rmm}:\;0\leq \ttt\leq k\beta\big]>|x|_{\rmm}\Big)\\
\qquad\qquad\leq\diy\frac{2}{|x|_{\rmm}+\sqrt{|x|^2_{\rmm}+4/\pi}}
\;\frac{e^{-|x|_{\rmm}/\beta}}{2\pi (1-\orho )}.
\end{array}\eqno (4.16)$$
Consequently, the contribution of this sum to $\diy\int_{\bbR^2} \rd x$ also does not exceed
a constant.

\medskip

More generally, for a given $r>R(L)$ we consider the contribution into (4.14) from loops $\om^*$
with $x(\om^*)\not\in\Lam_r$ such that $|\om^*(\ttt )|_{\rmm}=r$ for some $\ttt\in [0,k(\om^*)\beta ]$.
Repeating the above argument, we conclude that this contribution again is less than or equal to
a constant times $\tau_L(r)$. Note that all constants can be made uniform; this implies that
$$\begin{array}{l}(4.14)\diy\leq\frac{C_0}{\left[Q(L-R(L)-\eps-a/2)\right]^2}\\
\qquad\diy\times \bigg[
\big(\log\,\log\;L\big)^{3/2}
+ \int_{R(L)}^L\left[q\left(r-R(L)-\eps-a/2\right)\right]^2\rd r\bigg]\,.\end{array}
\eqno (4.17)$$
As in \cite{R2}, the quantity in the RHS of (4.17) (which is $=C_0\gamma (L)$)
goes to $0$ as $L\to\infty$. This finishes the proof of Proposition 4.1.

\medskip

It is instructive to note that the relation (4.9) does not require a smallness for $\eps$.

We now pass to random variable $\Sigma^{(2)}_L=\Sigma^{(2,1)}_L+\Sigma^{(2,2)}_L$.

\medskip

\medskip

{\bf Proposition 4.2.} {\sl For $\eps $ small enough, }
$$\lim_{L\to\infty}\int_{\cW^*_\td(\bbR^2)}\mu (\rd\bOm^*)\Sigma^{(2)}_L(\bOm^*)
=0.\eqno (4.18)$$

\medskip

\medskip

{\it Proof of Proposition} 4.2. In the beginning, we again
use the Campbell theorem (in conjunction with an argument
similar to  Eqn (6.25) from \cite{R2}). Then the integral in (4.18)
is less than or equal to a constant (say, $C_1$) times the sum $I^{2,1}+I^{2,2}$. Here
the term $I^{2,1}=I^{2,1}_L$ is specified as follows:
$$\begin{array}{l}
I^{2,1}=\diy\int_0^\beta\rd\ttt\int\rd\om^*\sum_{0\leq l,l''<k(\om^*)}\bigg\{
{\mathbf 1}(\om^*(l\beta +\ttt )\leftrightarrow\om^*_1(l''\beta +\ttt ))\rho (\om^*)\\
\diy +\sum\limits_{m\geq 1}
\prod\limits_{1<i\leq m}\int\rd\om^*_i\sum\limits_{0\leq l_i,\ovl_i
<k(\om^*_i)}{\mathbf 1}(\om^*_{i-1}(\ovl_{i-1}\beta +\ttt )
\leftrightarrow\om^*_i(l_i\beta +\ttt ))\\
\quad\diy\times
{\mathbf 1}(\om^*_m(l_m\beta +\ttt )\leftrightarrow\om^*(l''\beta +\ttt ))
\rho(\om^*,\om^*_1,\ldots ,\om^*_m)\bigg\}\\
\qquad\quad\diy\times{\mathbf 1}\Big(|\om^*(l\beta +\ttt )|_{\rmm}\leq
|\om^*(l''\beta+\ttt )|_{\rmm}\Big)\\
\quad\diy\times\Big[\tau_L( |\om^*(l\beta +\ttt )|_{\rmm}-\eps -a/2)-\tau_L( |\om^*(l''\beta+\ttt )|_{\rmm})\Big]^2
\end{array}\eqno (4.19.1)$$
where the loop $\om^*_0$ has been identified as $\om^*$ and value $\ovl_0$ as $l$.

Likewise, the summand $I^{2,2}=I^{2,2}_L$ is given by
$$\begin{array}{l}
I^{2,2}=\diy\int_0^\beta\rd\ttt\int\rd\om^*\int\rd{\om^*}''\\
\quad\diy\times\sum\limits_{\substack{0\leq l<k(\om^*)\\ 0\leq l''<k({\om^*}'')}}
\bigg\{
{\mathbf 1}(\om^*(l\beta +\ttt )\leftrightarrow{\om^*}''(l''\beta +\ttt ))\rho (\om^*,{\om^*}'')\\
\quad\diy +
\sum\limits_{m\geq 1}\prod\limits_{1<i\leq m}\int\rd\om^*_i\sum\limits_{0\leq l_i,\ovl_i
<k(\om^*_i)}{\mathbf 1}(\om^*_{i-1}(\ovl_{i-1}\beta +\ttt )
\leftrightarrow\om^*_i(l_i\beta +\ttt ))\\
\quad\diy\times
{\mathbf 1}(\om^*_m(l_m\beta +\ttt )\leftrightarrow{\om^*}''(l''\beta +\ttt ))
\rho(\om^*,\om^*_1,\ldots ,\om^*_m,{\om^*}'')\bigg\}\\
\qquad\quad\diy\times{\mathbf 1}\Big(|\om^*(l\beta +\ttt )|_{\rmm}\leq |{\om^*}''
(l''\beta+\ttt )|_{\rmm}\Big)\\
\quad\diy\times\Big[\tau_L( |\om^*(l\beta +\ttt )|_{\rmm}-\eps -a/2)-\tau_L( |{\om^*}''(l''\beta+\ttt )
|_{\rmm})\Big]^2\end{array}\eqno (4.19.2)$$
where again the loop $\om^*_0$ has been identified as $\om^*$ and value $\ovl_0$ as $l$.

So, it suffices to verify that
$$\lim_{L\to\infty}I^{2,1}=\lim_{L\to\infty}I^{2,2}=0.\eqno $$
Both integrals are analysed in a similar fashion, and we focus on one of them,
say, $I^{2,2}$.

We use elementary bounds
$$\begin{array}{l}\diy
\Big[\tau_L(|\om^*(l\beta +\ttt)|_{\rmm}-\eps -a/2)
-\tau_L(|{\om^*}''_m(l''\beta +\ttt)|_{\rmm})\Big]^2\\
\quad\diy\leq \Big[|\om^*(l\beta +\ttt)|_{\rmm}-\eps -a/2-|{\om^*}''(l''\beta +\ttt))|_{\rmm}
\Big]^2Z_L(|\om^*(l\beta +\ttt )|_{\rmm})\end{array}\eqno (4.20.1)$$
with $Z_L$ given in (4.5), and
$$\begin{array}{l}\diy
\Big[|\om^*(l\beta +\ttt)|_{\rmm}-\eps -a/2-|{\om^*}''(l''\beta +\ttt))|_{\rmm}\Big]^2\leq 3(\eps +a/2)^2\\
\quad\diy + 3|\om^*(l\beta +\ttt)|^2_{\rmm} +3|{\om^*}''(l''\beta +\ttt))|^2_{\rmm}.
\end{array}\eqno (4.20.2)$$

Employing in addition the Ruelle bound, we conclude that (4.19.2) does not exceed
$$\begin{array}{l}
\diy3\int_0^\beta\rd\ttt\int\rd\om^*
\frac{\orho^{k(\om^*)}}{k(\om^*)}\int\rd{\om^*}''\frac{\orho^{k({\om^*}'')}}{k({\om^*}'')}\\
\quad\diy\times\sum\limits_{\substack{0\leq l<k(\om^*)\\ 0\leq l''<k({\om^*}'')}}
Z_L(|\om^*(l\beta +\ttt )|_{\rmm})\bigg\{
{\mathbf 1}(\om^*(l\beta +\ttt )\leftrightarrow{\om^*}''(l''\beta +\ttt ))\\
\qquad\diy
+\sum\limits_{m\geq 1}\prod\limits_{1<i\leq m}\int\rd\om^*_i\frac{\orho^{k(\om^*_i)}}{k(\om^*_i)}\\
\qquad\times\diy\sum\limits_{0\leq l_i,\ovl_i
<k(\om^*_i)}{\mathbf 1}(\om^*_{i-1}(\ovl_{i-1}\beta +\ttt )
\leftrightarrow\om^*_i(l_i\beta +\ttt ))\\
\quad\diy\times
{\mathbf 1}(\om^*_m(l_m\beta +\ttt )\leftrightarrow{\om^*}''(l''\beta +\ttt ))\bigg\}\\
\qquad\quad\diy\times{\mathbf 1}\Big(|\om^*(l\beta +\ttt )|_{\rmm}\leq |{\om^*}''(l''\beta+\ttt )|_{\rmm}\Big)\\
\quad\diy\times\Big[(\eps +a/2)^2
+ |\om^*_0(l_0\beta +\ttt)|^2_{\rmm} +|\om^*_m(l_m\beta +\ttt))|^2_{\rmm}
\Big]\,.\end{array}\eqno (4.21)$$

Expanding the sum of squares in the parentheses, we obtain three expressions;
in view of similarity of the argument used for analysing each of them, we focus
on the one with the term $ |\om^*(l\beta +\ttt)|^2_{\rmm}$:
$$\begin{array}{l}
\diy\int_0^\beta\rd\ttt\int\rd\om^*
\frac{\orho^{k(\om^*)}}{k(\om^*)}Z_L(|\om^*(l\beta +\ttt )|_{\rmm})\ \int\rd{\om^*}''
\frac{\orho^{k({\om^*}'')}}{k({\om^*}'')}\\
\qquad\diy
\times{\mathbf 1}\Big(|\om^*(l\beta +\ttt )|_{\rmm}\leq |{\om^*}''(l''\beta+\ttt )|_{\rmm}\Big)\\
\quad\diy\times\sum\limits_{\substack{0\leq l<k(\om^*)\\ 0\leq l''<k({\om^*}'')}}
|\om^* (l\beta +\ttt)|^2_{\rmm}\bigg\{
{\mathbf 1}(\om^*(l\beta +\ttt )\leftrightarrow{\om^*}''(l''\beta +\ttt ))\\
\qquad\diy
+\sum\limits_{m\geq 1}\prod\limits_{1<i\leq m}\int\rd\om^*_i\frac{\orho^{k(\om^*_i)}}{k(\om^*_i)}\\
\qquad\times\diy\sum\limits_{0\leq l_i,\ovl_i
<k(\om^*_i)}{\mathbf 1}(\om^*_{i-1}(\ovl_{i-1}\beta +\ttt )
\leftrightarrow\om^*_i(l_i\beta +\ttt ))\\
\quad\diy\times
{\mathbf 1}(\om^*_m(l_m\beta +\ttt )\leftrightarrow{\om^*}''(l''\beta +\ttt ))\bigg\}
.\end{array}\eqno (4.22)$$
Again, we can expand the curled brackets and will analyse the behavior of the technically most
involved sum that emerges:
$$\begin{array}{l}
\diy\int_0^\beta\rd\ttt\int\rd\om^*
\frac{\orho^{k(\om^*)}}{k(\om^*)}\int\rd{\om^*}''\frac{\orho^{k({\om^*}'')}}{k({\om^*}'')}\\
\quad\diy\times\sum\limits_{\substack{0\leq l<k(\om^*)\\ 0\leq l''<k({\om^*}'')}}
|\om^* (l\beta +\ttt)|^2_{\rmm} Z_L(|\om^*(l\beta +\ttt )|_{\rmm})\\
\qquad\diy\times\sum\limits_{m\geq 1}\prod\limits_{1<i\leq m}\int\rd\om^*_i\frac{\orho^{k(\om^*_i)}}{k(\om^*_i)}\\
\qquad\times\diy\sum\limits_{0\leq l_i,\ovl_i
<k(\om^*_i)}{\mathbf 1}(\om^*_{i-1}(\ovl_{i-1}\beta +\ttt )
\leftrightarrow\om^*_i(l_i\beta +\ttt ))\\
\quad\diy\times
{\mathbf 1}(\om^*_m(l_m\beta +\ttt )\leftrightarrow{\om^*}''(l''\beta +\ttt ))
.\end{array}\eqno (4.23)$$

The argument for estimating (4.23) starts with the analysis of the integral
$\diy\int\rd{\om^*}''\frac{\orho^{k({\om^*}'')}}{k({\om^*}'')}$ for fixed values of the
variables in the remaining integrals. To this end, we invoke the Fubini theorem
and properties of the Brownian bridge. This allows us to conclude that
$$\begin{array}{l}
\diy\int\rd{\om^*}''\frac{\orho^{k({\om^*}'')}}{k({\om^*}'')}\sum\limits_{0\leq l''<k({\om^*}'')}
{\mathbf 1}(\om^*_m(l_m\beta +\ttt )\leftrightarrow{\om^*}''(l''\beta +\ttt ))\leq\\
\quad
\diy\int_{\bbR^2}\rd x''\int_0^\beta \rd\ttt \int_{A[\om_{m-1}(\ttt ),\eps ]}\rd y\sum\limits_{k''\geq 1}
\orho^{k''}\frac{e^{-|y-x''|^2/(2\ttt)}}{2\pi \tt}
\frac{e^{-|y-x''|^2/(2(k''\beta -\ttt))}}{2\pi(k''\beta -\ttt)}\,.\end{array}\eqno (4.24)$$
Here
$$A[\om_{m-1}(\ttt ),\eps ]=\{y\in\bbR^2:a<|y-\om_{m-1}(\ttt )|<a+2\eps\}\eqno (4.25)$$
stands for an annulus of width $2\eps$ around the center $\om_{m-1}(\ttt )$. (Initially, point
$y\in A[\om_{m-1}(\ttt ),\eps ]$ emerges here as the point on the circle of radius
$|y-\om_{m-1}(\ttt )|$ about $\om_{m-1}(\ttt )$
where the loop $\om$ hits this circle while $t$ is the hitting time.)
The RHS of (4.24) yields a quantity $\leq C_2\eps$.

This argument can be iterated for the integrals $\diy\int\rd\om^*_i\frac{\orho^{k({\om^*_i}'')}}{k({\om^*_i}'')}$
where we have to take into account the double sum $\diy\sum\limits_{0\leq l_i,\ovl_i
<k(\om^*_i)}$.
However, it only affects the constant in front of $\eps$.

At the end, assuming that $\eps >0$ is small enough we arrive at the following bound for (4.23):
$$\diy \frac{C_3\eps}{1-C_4\eps}\int_0^\beta\rd\ttt\int\rd\om^*
\frac{\orho^{k(\om^*)}}{k(\om^*)} \sum\limits_{0\leq l<k(\om^*)}
|\om^* (l\beta +\ttt)|^2_{\rmm} Z_L(|\om^*(l\beta +\ttt )|_{\rmm}).\eqno (4.26)$$
The integral (4.26) is analysed in the same manner as in Proposition 4.1 (cf. (4.9))
and tends to $0$. (The presence of the sum $\diy\sum\limits_{0\leq l<k(\om^*)}$ in (4.26)
does not affect the core of the argument.)

\medskip

This completes the proof of Proposition 4.2 and Lemma 4.1.

\vskip 1 truecm

\centerline{\large{\bf 5. Estimates for the change in the energy.}}
\centerline{\large{\bf Concluding remarks}}

\vskip 1 truecm

In this section we assess the expression 
$$\exp\;\Big[
h(\tT^+_L(s)\Om^*_{\Lam }|\bOm^*_{\Lamc})
+h(\tT^-_L(s)\Om^*_{\Lam }|\bOm^*_{\Lamc})-
2h(\Om^*_{\Lam }|\bOm^*_{\Lamc})\Big]$$
(cf. (2.8)). The argument is based on the same
idea as in Sect 8.6 of \cite{R2} (again we partially borrow the system of notation
from there). In the course of the argument we will produce a further (and final) specification
of the set $\cG_L\subset\cW^*_\td(\bbR^2)$ of good LCs.
Namely, given $\bOm^*\in\cW^*_\td(\bbR^2)$, we set, as before,
$$\bOm^*_\Lam=\{\om^*\in\bOm^*:\;x(\bom^*)\in\Lam\},\;\;\;
\bOm^*_{\Lamc}=\{\om^*\in\bOm^*:\;x(\bom^*)\in\Lamc\}.$$
Then write
$$\begin{array}{l}
\diy h(\tT^+_L(s)\bOm^*_{\Lam }|\bOm^*_{\Lamc})
+h(\tT^-_L(s)\bOm^*_{\Lam }|\bOm^*_{\Lamc})-
2h(\bOm^*_{\Lam }|\bOm^*_{\Lamc})\\
\quad\diy=\int_0^\beta\rd\ttt
\Big\{E[\{\tT^+(s)\bOm^*_\Lam\}(\ttt )|\{\bOm^*_{\Lamc}\}(\ttt )]\\
\qquad\quad\diy +E[\{\tT^-(s)\bOm^*_\Lam\}(\ttt )|\{\bOm^*_{\Lamc}\}(\ttt )]
-2E[\{\bOm^*_\Lam\}(\ttt )|\{\bOm^*_{\Lamc}\}(\ttt )]\Big\}.
\end{array}\eqno (5.1)$$
Here $E\left[\{\tT^\pm (s)\bOm^*\}_\Lam(\ttt )|\{\bOm^*_{\Lamc}\}(\ttt )\right]$ is defined as the sum
$$\begin{array}{l}
\diy
\frac{1}{2}\sum\limits_{x,x'\in\{\bOm^*_\Lam\}(\ttt )}
V\left(|x\pm s{\wt t}(x)-x'\mp s{\wt t}(x')|\right)\\
\qquad\qquad\qquad\diy +\sum\limits_{\substack{x\in\{\bOm^*_\Lam\}(\ttt )\\
x'\in\{\bOm^*_{\Lamc}\}(\ttt )}}
V\left(|x\pm s{\wt t}(x)-x'\mp s{\wt t}(x')|\right)
\end{array}\eqno (5.2)$$
while $E[\{\bOm^*\}(\ttt )|\bOm^*_{\Lamc}(\ttt )]$ is obtained by omitting the terms
containing the shift-vector $s$. Cf. (2.3.11.\bfI)--(2.3.12.\bfI). Recall, our aim is to guarantee that
on the good set $\cG_L$, the absolute values of the variables $\Sigma^{(i)}_L(\bOm^*)$ are small. Two
straightforward bounds turn out to be helpful:
$$\begin{array}{l}
\Big|E[\{\tT^+(s)\bOm^*_\Lam\}(\ttt )|\{\bOm^*_{\Lamc}\}(\ttt )]
+E[\{\tT^-(s)\bOm^*_\Lam\}(\ttt )|\{\bOm^*_{\Lamc}\}(\ttt )]\\
\qquad\qquad\qquad\diy -2E[\{\bOm^*_\Lam\}(\ttt )|\{\bOm^*_{\Lamc}\}(\ttt )]\Big|
\leq {\ov V}^{(2)}\ts^2\\
\quad\diy\times\Bigg\{\frac{1}{2}\sum\limits_{x,x'\in\{\bOm^*_\Lam\}(\ttt )}
+\sum\limits_{\substack{x\in\{\bOm^*_\Lam\}(\ttt )\\
x'\in\{\bOm^*_{\Lamc}\}(\ttt )}}\Bigg\}
|{\wt t}(x)-{\wt t}(x')|^2{\mathbf 1}(|x-x'|\leq\rR_0)
\end{array}\eqno (5.3)$$
and
$$\begin{array}{l}
\diy\frac{1}{3}|{\wt t}(x)-{\wt t}(x')|^2\\
\quad\diy\leq |{\wt t}(x)-\tau_ L(|x|_{\rmm})|^2+|\tau_L(|x|_{\rmm})-\tau_L(|x'|_{\rmm})|^2
+|\tau_L(|x'|_{\rmm})-{\wt t}(x')|^2.\end{array}\eqno (5.4)$$
These bounds yield that
$$\begin{array}{c}
\diy \left|h(\tT^+_L(s)\bOm^*_{\Lam }|\bOm^*_{\Lamc})
+h(\tT^-_L(s)\bOm^*_{\Lam }|\bOm^*_{\Lamc})-
2h(\bOm^*_{\Lam }|\bOm^*_{\Lamc})\right|\\
\diy\leq 3 {\ov V}^{(2)}\ts^2\int_0^\beta\rd\ttt\Bigg\{\frac{1}{2}\sum\limits_{x,x'\in\{\bOm^*_\Lam\}(\ttt )}
+\sum\limits_{\substack{x\in\{\bOm^*_\Lam\}(\ttt )\\
x'\in\{\bOm^*_{\Lamc}\}(\ttt )}}\Bigg\}{\mathbf 1}(|x-x'|\leq\rR_0)\\
\diy\times\Big[2|{\wt t}(x)-\tau_ L(|x|_{\rmm})|^2+|\tau_L(|x|_{\rmm})-\tau_L(|x'|_{\rmm})|^2
\Big]\\
=:\Sigma^{(3)}_L(\bOm^*)
+\Sigma^{(4)}_L(\bOm^*)\end{array}\eqno (5.5)$$
where variables $\Sigma^{(3)}_L$ and $\Sigma^{(4)}_L$ emerge when we expand the
sum of squares in the parentheses.

As above, we will try to make sure that the expected values of variables $\Sigma^{(3)}_L$ and
$\Sigma^{(4)}_L$ vanish as $L\to\infty$:

\medskip

\medskip

{\bf Lemma 5.1.} {\sl
$$\diy\lim_{L\to\infty}\int\mu (\rd\bOm^*)
\Sigma^{(3)}_L(\bOm^*)=\lim_{L\to\infty}\int\mu (\rd\bOm^*)
\Sigma^{(4)}_L(\bOm^*)=0.\eqno (5.6)$$}

\medskip

\medskip

{\it Proof of Lemma} 5.1. As before, we focus on one of the relations in Eqn (5.6),
say, for $\Sigma^{(4)}_L$. It is instructive to expand
$$\Sigma^{(4)}_L(\bOm^*)=\Sigma^{(4,1)}_L(\bOm^*)+\Sigma^{(4,2)}_L(\bOm^*).$$
Here  $\Sigma^{(4,1)}_L(\bOm^*)$ gives a single-loop contribution to
$\Sigma^{(4)}_L(\bOm^*)$ while $\Sigma^{(4,2)}_L(\bOm^*)$ yields a contribution
from pairs of loops:
$$\begin{array}{l}
\diy\Sigma^{(4,1)}_L(\bOm^*) = \sum\limits_{\om^*\in\bOm^*_\Lam}\int_0^\beta\rd\ttt
\Bigg\{\sum\limits_{0\leq l<{\ov l}<k(\om^*)}\\
\qquad\diy\times\Big[\tau_L\big(\big|\om^*(l\beta +\ttt)\big|_{\rmm}\big)
-\tau_L\big(\big|\om^*({\ov l}\beta +\ttt)\big|_{\rmm}\big)
\Big]^2\\
\qquad\qquad\diy\times{\mathbf 1}\Big(\big|\om^*(l\beta +\ttt)-|\om^*({\ov l}\beta +\ttt \big|<\rR_0 \Big)\\
\quad \diy +\sum\limits_{{\om^*}'\in\bOm^*_{\Lamc}}\sum\limits_{\substack{0\leq l<k(\om^*)\\
0\leq l'<k({\om^*}')}}\Big[\tau_L\big(\big|\om^*(l\beta +\ttt)\big|_{\rmm}\big)
-\tau_L\big(\big|{\om^*}'(l'\beta +\ttt)\big|_{\rmm}\big)\Big]^2\\
\qquad\qquad\diy\times{\mathbf 1}\Big(\big|\om^*(l\beta +\ttt)-|{\om^*}'(l'\beta +\ttt \big|<\rR_0 \Big)\Bigg\}
\end{array}\eqno (5.7.1)$$
and
$$\begin{array}{l}
\Sigma^{(4,2)}_L(\bOm^*)=
\diy \frac{1}{2}\int_0^\beta\rd\ttt\sum\limits_{\substack{\om^*,{\om^*}'\in\bOm^*_\Lam\\ \om^*
\neq{\om^*}'}}\;
\sum\limits_{\substack{0\leq l<k(\om^*)\\ 0\leq l'<k({\om^*}')}}\\
\qquad\diy\times\Big[\tau_L\big(\big|\om^*(l\beta +\ttt)\big|_{\rmm}\big)
-\tau_L\big(\big|{\om^*}'(l'\beta +\ttt)\big|_{\rmm}\big)\Big]^2\\
\qquad\qquad\diy\times{\mathbf 1}\Big(\big|\om^*(l\beta +\ttt)-|{\om^*}'(l'\beta +\ttt \big|
<\rR_0 \Big).\end{array}\eqno (5.7.2)$$
(The factor $3 {\ov V}^{(2)}\ts^2$ carried from (5.5) has been discarded.)

Following Eqn (6.22) from \cite{R2}, we estimate: (a) when $\big|\om^*(l\beta +\ttt)\big|_{\rmm}\leq
\big|{\om^*}'(l'\beta +\ttt)\big|_{\rmm}$,
$$\begin{array}{l}
\Big[\tau_L\big(\big|\om^*(l\beta +\ttt)\big|_{\rmm}\big)
-\tau_L\big(\big|{\om^*}'(l'\beta +\ttt)\big|_{\rmm}\big)\Big]^2\\
\leq \Big[ \big|\om^*(l\beta +\ttt)\big|_{\rmm}
-\big|{\om^*}'(l'\beta +\ttt)\big|_{\rmm} -\eps-a/2\Big]^2Z_L(|\om^*(l\beta +\ttt )|_{\rmm})
\end{array}$$
and (b) when $\big|{\om^*}'(l'\beta +\ttt)\big|_{\rmm}\leq
\big|\om^*(l\beta +\ttt)\big|_{\rmm}$,
$$\begin{array}{l}
\Big[\tau_L\big(\big|\om^*(l\beta +\ttt)\big|_{\rmm}\big)
-\tau_L\big(\big|{\om^*}'(l'\beta +\ttt)\big|_{\rmm}\big)\Big]^2\\
\leq \Big[ \big|{\om^*}'(l'\beta +\ttt)\big|_{\rmm}
-\big|\om^*(l\beta +\ttt)\big|_{\rmm} -\eps-a/2\Big]^2Z_L(|{\om^*}'(l\beta +\ttt )|_{\rmm})
\end{array}$$
where $Z_L$ has been defined in (4.5).

After substituting these estimates in $(5.1)$, the relation $\diy\int\mu (\rd\bOm^*)
\Sigma^{(4)}_L(\bOm^*)$ $\to 0$ is verified in the same way as in
Proposition 4.1. This completes the proof of Lemma 5.1.

\medskip

Lemma 5.1 (and the comments on other terms emerging from the bound (5.5)),
together with Lemmas 3.1 and 4.1, allows us to define the set $\cG_L$.
Namely,
$$\cG_L=\Big\{\bOm^*\in\cL_L:\;\Sigma^{(i)}_L(\bOm^*)<c,\;1\leq i\leq 4\Big\}
\eqno (5.8)$$
where $c\in (0,\infty )$ is a chosen constant (viz., $c=1/2$).
Applying the Chebyshev inequality guarantees

\medskip

\medskip

{\bf Lemma 5.2.} {\sl $\forall$ $\delta\in (0,1)$ and $c\in (0,\infty )$, there exists
$L^*_1\in (0,\infty )$ such that for $L>L^*_1$ the probability $\mu (\cG_L)\geq 1-\delta$. }

\medskip

\medskip

A formal summary of properties of transformations $\tT^\pm (s)$ is given in Theorem 5.1:

\medskip

\medskip

{\bf Theorem 5.1.} {\sl Given $\bOm^*\in\cG_L$, the transformations
$\tT^\pm_L (s):\;\bOm^*\mapsto{\wt\bOm}^*$ $\in\cW^*_\td(\bbR^2)$ possess the following properties:

{\rm{(i)}} The maps $\tT^\pm(s)$ are measurable and $1-1$.

{\rm{(ii)}} ${\wt\bOm}^*_{\Lamc}=\bOm^*_{\Lamc}$ and ${\wt\bOm}^*_{\Lam_0}=\tS (s)\bOm^*_{\Lam_0}$.
Moreover, there exists a $1-1$ correspondence between the loops ${\wt\om}^*\in{\wt\bOm}^*_\Lam$
and $\om^*\in\bOm^*_\Lam$ such that ${\wt\om^*}$ is obtained as a deformation of $\om^*$
via tuned shifts of $\ttt$-sections, in the manner
described in Section $3$. In particular, $k({\wt\om}^*)=k(\om^*)$.

{\rm{(iii)}} The equality {\rm{(2.6)}} holds true, where the expression \\
$\Big[J^+_L(\Om^*_\Lam\vee\bOm^*_{\Lamc})J^-_L(\Om^*_\Lam\vee\bOm^*_{\Lamc})\Big]^{1/2}$
is close to $1$ uniformly in $\bOm^*$ for $L$ large.

{\rm{(iv)}} The quantity {\rm{(2.8)}} is close to $1$ uniformly in $\bOm^*$ when  $L$ is large enough.
}

\medskip

The assertion of Theorem 2.1 then follows.

\vskip 2 truecm

\subsection*{Acknowledgments}
This work has been conducted under Grant 2011/20133-0 provided by
the FAPESP, Grant 2011.5.764.35.0 provided by The Reitoria of the
Universidade de S\~{a}o Paulo, Grants 2012/04372-7 and
11/51845-5 provided by the FAPESP. The authors
express their gratitude to NUMEC and IME, Universidade de S\~{a}o Paulo,
Brazil, for the warm hospitality.

\end{document}